\documentclass[preprint,showpacs,aps,floatfix]{revtex4}
\usepackage{graphicx}
\usepackage{dcolumn}
\usepackage{bm}

\def\by#1#2{{\displaystyle {#1}\over \displaystyle {#2}}}
\def\d{{\rm d}}
\preprint {IMSc/2009/08/11}
\begin{document}
\title{Effect of tau neutrino contribution to muon signals at neutrino
factories}
\author{D. Indumathi and Nita Sinha}

\affiliation
{The Institute of Mathematical Sciences, Chennai 600 113, India.\\
}
\date{\today} 

\begin{abstract}
  We discuss precision measurements of the leading atmospheric
  parameters at a standard neutrino factory with a detector that is
  sensitive to muons alone. The oscillation of the muon- and
  electron neutrinos in the neutrino factory beam to tau neutrinos adds to
  the muon events sample (both right sign and wrong sign) via leptonic
  decays of the taus produced through charge-current interactions
  in the detector. In particular, we study how this affects a precision
  measurement of the atmospheric mixing parameters and the deviation of
  $\nu_\mu\leftrightarrow \nu_\tau$ mixing from maximality. In spite of
  the enhancement of the number of events due to the additional tau
  contribution, the determination of the atmospheric mixing angle and the
  deviation from maximality will be poorer. We show that it is impossible
  to devise satisfactory cuts to remove this tau contamination. Neglect
  of these tau contributions will lead to an incorrect conclusion about
  the precision obtainable at a neutrino factory.

\end{abstract}


\maketitle

\section{Introduction}

Neutrino oscillations \cite{theory} are now well-established. The
relevant mass-squared differences and mixing angles have also been
studied in various experiments \cite{solar,atm,reactor,ktok,minosnew};
the study of neutrino oscillation physics is now at the precision stage
with many existing and proposed near-future experiments. A natural
progression in the precision study is to probe oscillation physics with
neutrino factory beams.

In a neutrino factory, a $\mu^+$ or $\mu^-$ beam from a storage ring
decays down a long straight section of the beam-line that points at a
far-end detector. The decay of the muons produces a spectrum of electron
and muon neutrinos via $\mu^+ \to \overline{\nu}_\mu e^+ \nu_e$ and $\mu^-
\to \nu_\mu e^- \overline{\nu}_e$, that is extremely well-understood.
Charged-current (CC) interactions of the muon anti-neutrinos (neutrinos)
in the detector lead to the production of $\mu^+$ ($\mu^-$) leptons. The
electron neutrinos (anti-neutrinos), on the other hand, can {\it oscillate}
into muon neutrinos (anti-neutrinos) which in CC interactions in the
detector, produce $\mu^-$ ($\mu^+$) leptons, or muons with charge
opposite to that from the unoscillated case and hence get
detected as ``wrong sign'' (WS) muons.

In general, the focus of studies \cite{wrongsign,IDS-ISS} with neutrino
factories (and the many papers and effort towards specifying the
parameters of such a future neutrino factory) has been to pin down the
as-yet unknown neutrino oscillation parameters, in particular, the
across-generation 13 mixing angle $\theta_{13}$ and the CP violating
phase $\delta_{CP}$, in addition to other possibilities such as the
mass hierarchy (or mass ordering in the 23 sector) and the octant of
the 23 mixing angle $\theta_{23}$ (whether it is larger or smaller
than $\pi/4$). All these can be best studied through measuring the WS
events, as discussed above, in detectors capable of charge
identification. The so-called right-sign (RS) (having the same charge as
unoscillated) muon events are useful for precision measurements of the
leading atmospheric parameters $\theta_{23}$ and the atmospheric mass
squared difference, $\Delta m^2$, apart from their use in understanding
cross-section and flux uncertainties. In particular, they are sensitive
probes of whether $\nu_\mu\leftrightarrow \nu_\tau$ mixing is maximal,
when $\theta_{23} = \pi/4$; we shall refer to this value of $\theta_{23}$
as being maximal. Such maximal mixing usually arises from an underlying
symmetry \cite{symm} that would dictate the structure of the neutrino
mixing matrix. Measurement of deviation from maximality is therefore of
great significance in developing models for neutrino masses and mixings.

In this paper, we address the little-studied issue of contamination of
the (right or wrong sign) muon events sample from
oscillations of the muon or electron neutrinos or anti-neutrinos to tau
neutrinos or anti-neutrinos, which, through CC interactions in the
detector, result in tau leptons that subsequently decay to muons.
While tau-production in neutrino-nucleus collisions
has been studied in great detail, for example in Ref.~\cite{yokoya},
there are not many studies on the implications of the tau
contamination (through leptonic decay) of the CC muon or electron
events.

Tau contributions naturally arise in multi-GeV CC neutrino--nucleus
interactions. Although the tau CC events are kinematically suppressed due
to the large tau mass, $m_\tau = 1.78$ GeV, the total contribution from
taus is substantial since $\nu_\mu\leftrightarrow\nu_\tau$ oscillations
are large, being driven by a nearly maximal $\theta_{23}$. Here we focus
on how this contribution affects a precision measurement of the
deviation of $\nu_\mu\leftrightarrow\nu_\tau$ mixing from maximal, i.e.,
a precision measurement of $\theta_{23}$ and its deviation from maximal.

Tau contributions enhance both the RS and WS event rates in both the muon
and electron sectors, since the leptonic tau decay fraction into both
electrons and muons is about the same ($\approx 17\%$). In particular, we
find that tau contributions substantially enhance the {\em right-sign}
muon event rates, especially at small final lepton energies, {\em
independent of $\theta_{13}$}. However, due to a different parameter
dependence, the tau contribution to muons {\em worsens} the determination
of the deviation from maximality of $\theta_{23}$ while leaving the
determination of $\vert \Delta m^2 \vert$ unchanged. We also show that it
is practically impossible to devise satisfactory cuts to remove this tau
contamination. Hence, studies (beyond those discussed here) at neutrino
factories must first bring the relatively little-studied uncertainties
due to tau production and decay under control in order to
achieve the expected precision measurements with muon detection.
 
The paper is organized as follows. In the next section, we briefly
describe the physics of neutrino oscillations as is relevant for this
study. We also list here the current limits on the oscillation
parameters. In Section 3, we discuss the inputs such as the neutrino
fluxes and cross-sections and preferred base-lengths. While we use
standard packages for the deep inelastic scattering (DIS)
cross-sections involving muon neutrinos, we use the mass-corrected
leading order (LO) expressions for the tau contributions as discussed
in Ref.~\cite{yokoya}, as well as comment on the next-to-leading order
(NLO) contributions. In Section 4, we calculate numerically the event
rates for various processes at an iron calorimeter detector capable of
detecting muons with charge identification, such as
the proposed ICAL/INO or the numerically mooted MIND. The numerical
computation involves the calculation of the neutrino factory fluxes,
the probability amplitudes using a solver for the neutrino evolution
equation, which has been described in Ref.~\cite{octant}, calculation
of the double differential cross section and in addition, the
differential decay rate of the taus to muons, in case of the tau
contribution. We conclude in Section 5.

\section{Oscillation probabilities}
With a long baseline, the large matter effects in the course of
neutrino propagation help in improved sensitivity to many parameters.
In general, the 3-flavor probabilities, $P_{ij}$, of flavor $\nu_i$
oscillating into flavor $\nu_j$, depend on all the oscillation
parameters (in vacuum): the mixing angles $\theta_{ij}, i,j=1,2,3$,
the CP violating phase, $\delta_{CP}$ and the mass-squared differences
$\Delta_{ij}= m_i^2-m_j^2$, as well as the density of Earth's matter.
The following analytic expressions \cite{matprob} for these probabilities
for neutrino propagation in earth matter in the constant density
approximation, are useful to qualitatively exhibit the parameter dependence
and for understanding the exact numerical results of the next section:
\begin{eqnarray}
\nonumber \label{eq:p3num}
P_{e\mu} & \approx & \sin^2 \theta_{23} \sin^2 2\theta_{13}^{m}\,
\sin^2 \Delta_{31}^m~,\nonumber \\
P^m_{e\tau} & \approx & \cos^2 \theta_{23} \sin^2 2\theta_{13}^{m}\,
\sin^2 \Delta_{31}^m~,\nonumber \\
P^m_{\mu\mu} & \approx & 1 - \sin^2
2\theta_{23} \, \left[\sin^2\theta_{13}^{m} \sin^2\Delta_{21}^m +
\cos^2\theta_{13}^{m}\sin^2\Delta_{32}^m \right]
- \sin^4\theta_{23} \sin^2 2\theta_{13}^{m} \sin^2\Delta_{31}^m~,\nonumber \\
P^m_{\mu\tau} & \approx & \sin^22\theta_{23}\,\left[ \cos^2
\theta_{13}^{m} \sin^2\Delta_{32}^m -\cos^2 \theta_{13}^{m}
\sin^2 \theta_{13}^{m} \sin^2\Delta_{31}^m + \sin^2 \theta_{13}^{m}
\sin^2\Delta_{21}^m \right]~,
\end{eqnarray}
where terms involving solar mass terms have been ignored \cite{IM} and
the superscript $m$ refers to mixing angles and mass square differences
in matter.

For the atmospheric mass squared difference we use the notation
\cite{foglipar} $\Delta m^2 \equiv m_3^2-\frac{1}{2} (m_1^2+m_2^2)$,
so that a normal or inverted hierarchy is simply indicated by a sign
(and not magnitude) change in this parameter.

We use a Runge-Kutta solver to calculate the oscillation probabilities
for various energies and path lengths. Some technical details are given
in Ref.~\cite{octant}. All numerical results presented in this paper
have been obtained using the density profile of the Earth as given by
the Preliminary Reference Earth Model ({\sc prem}) \cite{prem} and
numerically evolving the flavor eigenstates through Earth's matter. In
particular, the approximate expressions for the probabilities as shown
in Eq.~(\ref{eq:p3num}) are not used.

\subsection{Existing constraints on the oscillation parameters}

The currently accepted \cite{schwetz} best-fit values of the
oscillation parameters are summarized in Table~\ref{tab:bestfit}. The
sign of the mass-squared difference $\Delta m^2$ as well as any
possible deviation of $\theta_{23}$ from maximality (as well as its
octant) are not yet determined. Also, there exists just an upper bound
on the effective (13) mixing angle,
while the CP violating phase $\delta_{CP}$ is unknown.

Here, we will focus on precision measurement of deviation of
$\theta_{23}$ from maximality, i.e., deviation from $\pi/4$. We base the
detector near the magic base-line, at $L=7,400$ km, so that the results
are insensitive to the unknown $\delta_{CP}$. We
will assume that a combination of solar neutrino experiments and the
KamLAND reactor\cite{solar,reactor} experiment completely determines
the parameters $\Delta_{21}$ and $\theta_{12}$. Since these play an
insignificant role in determination of the atmospheric parameters,
we set them to the best-fit values given in Table~\ref{tab:bestfit}
and do not vary them in this analysis.

\begin{table}
\begin{tabular}{|l|r|c|} \hline
Parameter & \multicolumn{1}{c|}{Best-fit value} & 2-$\sigma$ error \\ \hline
$\Delta_{21} [10^{-5}]$ eV$^2$ & $7.65^{+0.23}_{-0.20}$ & 7.25--8.11 \\ \hline
$|\Delta m^2|$ [10$^{-3}]$ eV$^2$ & $2.40^{+0.12}_{-0.11}$ & 2.18--2.64\\ \hline
$\sin^2 \theta_{12}$; \quad $\left[\theta_{12} \right]$
 & $0.304^{+0.022}_{-0.016}$; \quad $\left[33.5^\circ \right]$ & 0.27--0.35 \\
$\sin^2 \theta_{13}$; \quad $\left[\theta_{13} \right]$ &
$0.01^{+0.016}_{-0.011}$ & $< 0.040$; \quad $\left[11.5^\circ\right]$ \\
$\sin^2 \theta_{23}$; \quad $\left[\theta_{23} \right]$
 & $0.50^{+0.07}_{-0.06}$; \quad $\left[ 45.0^\circ \right]$ &
 0.39--0.63 \\ \hline
\end{tabular}
\caption{Table showing currently accepted \cite{schwetz} best-fit values
of oscillation parameters with 2$\sigma$ errors. In the case of the
mixing angle $\theta_{13}$, a 2$\sigma$ upper bound is shown.}
\label{tab:bestfit}
\end{table}

\subsection{Tau contributions and $\theta_{23}$} \label{sec:tau} The
$\nu_\mu \leftrightarrow \nu_\tau$ oscillations are large, being
driven by a nearly maximal $\theta_{23}$. Furthermore, as seen from
Eq.~(\ref{eq:p3num}), $P_{\mu\tau}$ remains large even if $\theta_{13}$
vanishes. Although there is kinematic suppression of the CC cross-section
for tau production due to the large tau mass and the decay rate into
electrons and muons is about 17\%, there is still a sizeable tau
production rate, which feeds into the muon production rate upon
leptonic tau decay. In particular, this adds to the {\em right sign}
muon event rate, and can even double it at low muon energies, since the tau
decays preferentially produce low energy muons. It should therefore be
taken into account in a precision analysis.

On the other hand, as seen from the same equation, $P_{e\tau}$ is
proportional to powers of $\sin\theta_{13}$ and the tau contribution to
WS muons is therefore small. It is of course comparable to the direct (from
$\nu_e$) wrong sign muon production rate to which it adds, although
rather smaller since the large tau mass suppresses the cross-section;
in addition, the contributing events are from further decay of the tau
into muons and hence the rate is further suppressed. While the analysis
of wrong sign muon events to determine parameters such as $\theta_{13}$,
CP phase, etc., should perhaps include this contribution, it will
have a negligible effect on the analysis at hand, viz., deviation of
$\theta_{23}$ from maximality; however,in our numerical analysis we do
include this contribution.

In short, tau contribution to muon events from $\nu_\mu \leftrightarrow
\nu_\tau$ oscillations is large and adds to the right sign events; its
contribution from $\nu_e \leftrightarrow \nu_\tau$ oscillations is small
and adds to the wrong sign events. Hence there are additional muon events
from $\nu_\tau$ in both RS and WS sectors to the direct events arising
from (oscillated or unoscillated) $\nu_\mu$ interactions.

\paragraph{RS Events}: The RS events are not very sensitive to the
octant of $\theta_{23}$. Notice that two of the terms of $P_{\mu\mu}$
depend only on $2\theta_{23}$ and are insensitive to the octant of
$\theta_{23}$. Further, for small $\theta_{13}$, the octant dependent
term in $P_{\mu\mu}$ has negligible contribution, while $P_{\mu\tau}$
is completely independent of the octant. Similarly, individually,
both are sensitive to the {\em deviation} of $\theta_{23}$ from
maximality, but they have different dependences on this parameter.

To leading order in $\theta_{13}$, $P_{\mu\mu} \sim 1 -
\sin^22\theta_{23}\cdot F$ while $P_{\mu\tau} \sim
\sin^22\theta_{23}\cdot F$, where $F = \sin^2\Delta_{32}^m$. Since the
$\theta_{23}$ dependent terms come with opposite sign, the combination
of muons from direct production and from tau decays marginally decreases
the sensitivity of the event rates to this angle (and its deviation from
maximality). However, as we shall see later, it is almost impossible to
remove the tau events. Any cuts that attempt to do so drastically reduce
the direct muon events as well and hence worsen the sensitivity to the
oscillation parameters. Thus, the combined contribution of events from
direct muon production and those from tau decay is less sensitive to the
deviation from maximality. Neglect of the tau contribution will lead to
an incorrect conclusion about the precision possible for the deviation
from maximality. This is the crux of our result.

\paragraph{WS Events}: Notice that $P_{e\mu}$ and $P_{e\tau}$ are both
sensitive to the octant of the 23 mixing angle, $\theta_{23}$, while
their sum is less sensitive (indeed, the sum would be wholly
independent of $\theta_{23}$, if one were to neglect the differences
in the $\nu_\mu$-nucleon and $\nu_\tau$-nucleon cross-sections). Hence
the tau contribution affects precision measurements in the WS sector
as well.

Here we have focused on events with muons in the final state. However,
taus also equally contribute to events with electrons in the final
state. These events are interesting for other precision measurements
and have been discussed in the context of $\nu_\mu\to\nu_e$ appearance
or the ``platinum channel'' where charge identification of electrons
is also contemplated. Here the situation with respect to the tau
contribution is drastically different. The tau contribution to
electron events from $\nu_\mu \leftrightarrow \nu_\tau$ oscillations
(with subsequent decay of tau into electrons) is large and adds to the
wrong sign events. Hence signatures in the ``platinum channel'' may
suffer from these large additional contributions, independent of
$\theta_{13}$. This platinum channel is well-motivated when
$\theta_{13}$ is large (when the ``direct'', that is, not
tau-generated, contribution is large); in such a case, the tau
contamination from $\nu_e \leftrightarrow \nu_\tau$ oscillations also
turns on and becomes a small addition to the right sign electron events.

Hence the inclusion of tau leptonic decay boosts the signal for muon RS
events and spoils the purity of the electron WS events. Details of this
contribution to the electron events will be presented separately.

\section{The Inputs: Fluxes, kinematics, cross-sections}

\subsection{The neutrino factory fluxes}

We assume a basic muon storage ring configuration \cite{IDS-NF-base}
with muon beam energy $E_b = 25$ GeV, with $5 \times 10^{20}$ useful
decays per year. The muon antineutrino (neutrino) and electron neutrino
(antineutrino) distributions for $\mu^\pm$ decay in the muon rest frame
are given by,
\begin{eqnarray}
\by{\d^2N_{\bar{\nu}_\mu,\nu_\mu}}{\d x\d\Omega} & \propto &
\by{2x^2}{4\pi} \left[(3-2x)\mp (1-2x)P_\mu \cos\theta\right]~, \nonumber \\
\by{\d^2N_{\nu_e,\bar{\nu}_e}}{\d x\d\Omega} & \propto &
\by{12x^2}{4\pi} \left[(1-x)\mp (1-x)P_\mu \cos\theta\right]~, \nonumber
\label{flux0}
\end{eqnarray}
where $P_\mu$ is the average muon polarization along the beam
direction, $E_\nu$ denotes the neutrino energy, $x = 2E_\nu/m_\mu$ and
$\theta$ is the angle between the neutrino momentum and muon spin vector.

For unpolarized muon beams, in the laboratory frame the boosted flux
distributions are given by,
\begin{eqnarray}
\by{\d^2N_{\nu_\mu}}{\d y\d\Omega_{\rm lab}} & = & \by{4n_\mu}{\pi L^2
m_\mu^6} E_b^4y^2 \left(1-\beta\cos\varphi\right)\, 
\left[3m_\mu^2-4E_b^2 y \left(1-\beta\cos\varphi\right)\right]~, \nonumber \\
\by{\d^2N_{\nu_e}}{\d y\d\Omega_{\rm lab}} & = & \by{24n_\mu}{\pi L^2 
m_\mu^6} E_b^4y^2 \left(1-\beta\cos\varphi\right)\,
\left[m_\mu^2-2E_b^2 y \left(1-\beta\cos\varphi\right)\right]~, \nonumber
\label{eq:flux}
\end{eqnarray}
where $y=E_\nu/E_b$, $n_\mu$ is the number of useful muons per year,
$L$ is the base-line distance from source to detector, and $\varphi$ is the
angle between the beam axis and the direction to the detector (which is
assumed to be the forward or $z$-direction).

In general, the decaying muon beam has a finite divergence ($\sim mr$
or less) so that it has an angular spread about the central beam
direction. Further, the neutrinos are emitted in a forward cone that
becomes narrower as $E_b$ increases. We can relate the
neutrino--muon opening angle $\varphi$ to the angles $\alpha$ (muon
direction with respect to the $z$-axis) and $(\theta', \phi')$
(neutrino direction with respect to the $z$-axis) as,
$$
\cos\varphi = \cos\alpha\cos\theta' + \sin\alpha\sin\theta'\cos\phi'~.
$$
The neutrino fluxes can be (analytically) integrated over the muon
beam angle $\alpha$, assuming a gaussian angular divergence of the
muon beam around the $z$-axis with standard deviation taken to be
\cite{broncano} $\sigma = 0.1/\gamma$, with the usual definition of
$\gamma = E_b/m_\mu$. The resulting neutrino flux is a function of
$(\theta', \phi')$ alone. A trivial integration over $\phi'$ thus
gives the neutrino fluxes as a function of the neutrino angle,
$\theta'$. The kinematic boost ensures that about half of the
neutrinos are emitted within a cone $\theta' \le 1/\gamma$. Even
though the opening angle is very small, notice that the flux is not
uniform in $\theta'$; this can be important especially to understand
the fluxes at near detectors as well as its spread at the detector. We
average the forward flux over a small angle $\theta' < \epsilon =
0.3\sigma$ where the intensity is fairly flat as a function of
distance from the central axis to obtain the energy spectrum $\d N/\d
E_\nu$ of the electron and muon neutrinos. This corresponds to
knowing the neutrino beam direction to a precision of roughly 0.1 mr.

\begin{figure}[htp]
\includegraphics[width=\textwidth]{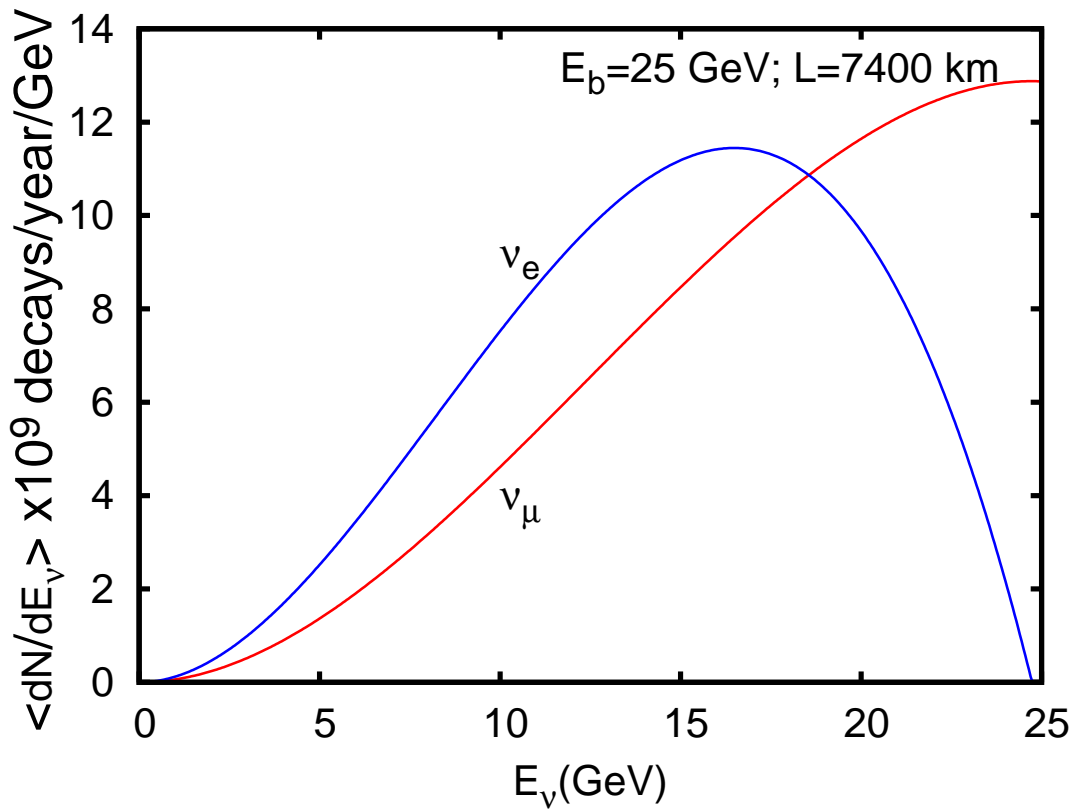}
\caption{Unoscillated electron and mu neutrino spectrum from a muon
factory at a distance of $L=7,400$ km, with $E_b=25$ GeV muons.}
\label{fig:spec}
\end{figure}

The resulting neutrino spectrum with gaussian angular spread averaged
over $(\theta' < \epsilon, 0 \le \phi' \le 2\pi)$ is
\begin{eqnarray}
\frac{\d N_\mu}{\d
E_\nu} \equiv \frac{1}{E_b}\frac{1}{\int\!\d\Omega} \, \int \d\Omega
\left\langle \by{\d N_\mu}{\d y\d\Omega}
\right\rangle_{\!G} & = & \by{4n_\mu \gamma^4 y^2}{\pi L^2 E_b} \left\{ 3
-4 y\gamma^2 -\by{\beta}{2}\left(3-8y\gamma^2\right) \left(1+c_\epsilon\right)
e^{-\sigma^2/2}\right. \nonumber \\
 & & \left. - \by{1}{3}y\left(\gamma^2-1\right)
\left[4+c_\epsilon+c_\epsilon^2+e^{-2\sigma^2}3c_\epsilon
\left(1+c_\epsilon\right)\right]\right\}~, \nonumber \\ 
\frac{\d N_e}{\d
E_\nu} \equiv \frac{1}{E_b}\frac{1}{\int\!\d\Omega} \,
\int \d\Omega \left\langle \by{\d N_e}{\d y\d\Omega}
\right\rangle_{\!G} & = & \by{24n_\mu \gamma^4 y^2}{\pi L^2 E_b} \left\{1
-2 y\gamma^2 -\by{\beta}{2}\left(1-4y\gamma^2\right) \left(1+c_\epsilon\right)
e^{-\sigma^2/2}\right.\nonumber \\
& & \left. - \by{1}{6}y\left(\gamma^2-1\right)
\left[4+c_\epsilon+c_\epsilon^2+e^{-2\sigma^2}3c_\epsilon
\left(1+c_\epsilon\right)\right]\right\}~,
\end{eqnarray}
where $c_\epsilon,s_\epsilon$ refer to $\cos\epsilon, \sin\epsilon$.
The (unoscillated) neutrino spectrum at a detector $L=7,400$ km away is
shown in Fig.~\ref{fig:spec}. The neutrino and anti-neutrino spectrum
of the same flavor (from muon beams of opposite charge) are the same
if the muon mass is neglected, as has been assumed here.

Note that, in the absence of oscillations, the number of events in the
neutrino sector is always a factor of two or more larger than in the case
of anti-neutrinos due to the larger neutrino cross-section. Hence, studies
of wrong-sign muons tend to focus on $\mu^+$ beams where the wrong sign
events correspond to $\mu^-$ from neutrino--nucleus interactions. Here,
we are interested in the right sign events, and so would prefer $\mu^-$
factory beams; however, we sum over events from both $\mu^-$ and $\mu^+$
beams, assuming an equal (and simultaneous) running time of 5 years in
each case.

\subsection{The kinematics}\label{sec:kinem}

We are interested in processes that yield muons (of either charge) in
the final state. These arise from CC interactions in which either muons
or taus are produced, with the latter decaying into muons. Studies with
neutrino factory beams typically use integrated cross-sections and aim at
reconstructing the initial neutrino energy and direction; here we focus
on the spectrum of the final state muon and hence require the detailed
kinematics of the interactions, both for direct muon production and via
tau decay. 

In the laboratory frame, a neutrino of flavor $l = \mu$ or $\tau$
and energy $E_\nu$ interacts with a nucleon of mass $M$ and produces
the corresponding charged lepton $l$ at an angle $\theta_l$ w.r.t. the
incident neutrino direction, which we take to be the $z$-axis. In the
case of $\nu_\tau$ interactions, the azimuthal angle of the muon from
tau decay, $\phi_\mu$, is also relevant: the tau is produced at a very
forward angle while $\phi_\mu$ is restricted by the decay kinematics. The
available phase space is restricted in both direct muon and tau production
due to the constraint on the available energy for the lepton: $E_-< E_l <
E_+$; see Appendix A for the detailed expressions on the constraints.

\begin{figure}[htp]
\includegraphics[width=\textwidth]{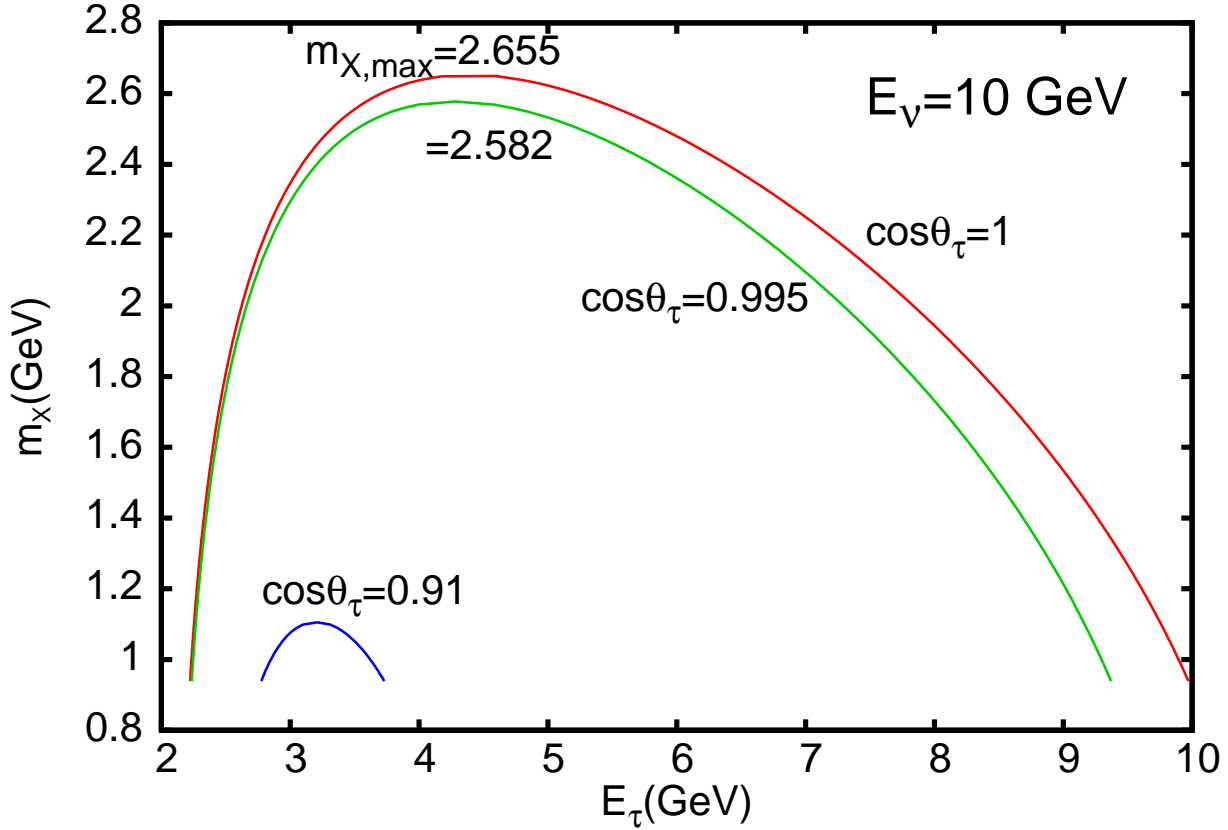}
\caption{Kinematics of $\nu_\tau$--nucleon CC interactions. The figure
shows the allowed parabolas of constant $\cos\theta_\tau$ in the
$m_X$-$E_\tau$ plane for $E_\nu=10$ GeV. The ends of the parabolas
(at $m_X = m_{X,{\rm min}}=M$) give the limits of the allowed tau lepton
energy, which are severely restricted for $\cos\theta_\tau$ away from 1.}
\label{fig:emX}
\end{figure}

The effect of this pinching in available energy, for the case of a $\tau$
lepton being produced, can be seen in Fig.~\ref{fig:emX} where the final
hadronic mass $m_X$ is plotted as a function of $E_\tau$. The notation
is standard: $m_X^2 = W^2 = (p+q)^2$, where $p$, $q$ are the nucleon
and the intermediate gauge boson 4-momenta in the laboratory frame.

The allowed tau energy limits $E_\pm$ for a given $\cos\theta_\tau$
correspond to the end points of the parabola in the $m_X$-$E_\tau$ plane.
For example, for a typical neutrino energy $E_\nu=10$ GeV,
there is hardly any allowed energy region for $\cos\theta_\tau=0.91$.
Hence, the tau leptons are produced in a very forward direction while the
direct muons, due to their lighter mass, are less restricted.

\subsection{The cross-sections}

Since the energies of interest range from a few GeV to 25 GeV, the CC
interactions include quasi-elastic (QE), resonance (Res) and deep
inelastic (DIS) processes.

Unlike many other calculations, we consider not the total but the double differential
cross-sections for muon production with detailed energy and direction
information so that we have better control on the spreads in the muon
energy and direction. Use of total cross-sections requires the
reconstruction of the original neutrino kinematics and introduces
larger errors due to large uncertainties in reconstruction of the
associated hadron kinematics. The differential cross-section may be
written in the form,
\begin{equation}
\by{\d\sigma}{\d E_l \d \cos\theta_l} = \by{G_F^2 \kappa^2}{2\pi}
\by{p_l}{M} \left\{ \sum_{i=1}^{5} a_i W_i\right\}~,
\label{eq:cross}
\end{equation}
where $G_F$ is the Fermi constant, $\kappa=M_W^2/(Q^2+M_W^2)$ is the
propagator factor with the $W$ boson mass, $M_W$, $p_l$ is the
magnitude of three-momentum of the charged lepton and the $W_i$ are
structure functions corresponding to the general decomposition of the
hadronic tensor for QE, Res and DIS processes. We have
\begin{eqnarray}
\sum_{i=1}^{5} a_i W_i & = & \left(2W_1+\by{m_l^2}{M^2} W_4\right)
\left(E_l-p_l\cos\theta_l\right) + W_2 \left(E_l +
p_l\cos\theta_l\right)~ \nonumber \\
 & & \pm \by{W_3}{M} \left(E_\nu E_l + p_l^2 - \left(E_\nu+E_l\right)
 p_l\cos\theta_l\right) - \by{m_l^2}{M} W_5~. 
\label{eq:Wi}
\end{eqnarray}
The detailed expressions for $W_i$ are taken from Ref.~\cite{yokoya}
where the specific structure functions are listed for QE, Res and
DIS leading order (LO) processes. A few comments are in order.

The $W_1$ and $W_2$ functions are those that appear in neutral current
(NC) lepton-nucleon scattering processes as well, while $W_3$ is a signature of
the parity violating nature of weak interactions. The structure
functions $W_4$ and $W_5$ only appear when the mass of the charged
lepton is taken into account. For consistency, we use non-zero masses
for both muon and tau leptons in the calculation; however, $W_4$ and
$W_5$ are actually relevant only for the heavy tau.

The quasi-elastic events dominate at lower energies. The separation
between Res and DIS is arbitrary and determined by a cut, chosen to be
$W > W_{\rm cut} = 1.4$ GeV.
This still leaves $Q^2 \equiv -q^2$ undetermined. However, reliable
perturbative estimates for DIS can only be made when $Q^2 >$ few
GeV$^2$. A few percent of the events satisfy the cut on $W$ for DIS
but have very low $Q^2$ (and hence very small values of the Bjorken
scaling variable, $x = Q^2/(2p\cdot q)$), thus making cross-section
estimates for these cases extremely uncertain. Removing these events
will lead to an underestimate of the actual number of DIS events;
estimating them by scaling their $Q^2$ to $Q^2_{\rm min} \sim 2$
GeV$^2$ overestimates the cross-sections since the structure functions
are large when $x$ is small.

For both mu and tau interactions we use the CTEQ6 LO
set of parton distribution functions. In this parameterization set, the
starting value is $Q_0 = 1.3$ GeV, or the charm mass. Since backwards
evolution is unstable, the problem lies with events with $Q^2 <
Q_0^2=1.69$ GeV$^2$.

\begin{figure}[hbp]
\includegraphics[width=\textwidth,bb=18 144 592 440, clip]{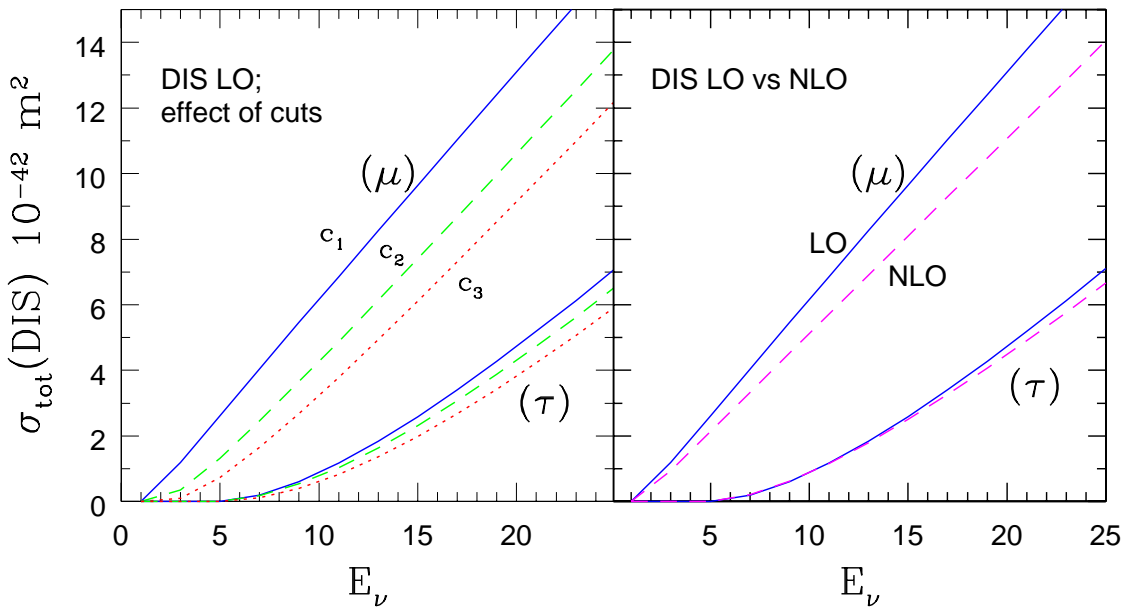}
\caption{Total deep inelastic scattering (DIS) CC $\nu_\tau$--nucleon
cross-section as a function of the neutrino energy. The left figure shows the
impact of different cuts on the cross-section. See text for details. The
right figure shows the effect of including NLO effects to the $c_1$
curve. The upper set of curves is for mu (or electron) DIS; the lower
ones for tau.}
\label{fig:dis}
\end{figure}

The effect of either choice (removing the event or adding it by estimating
its contribution at a larger $Q^2$) is shown in Fig.~\ref{fig:dis} for
both mu and tau interactions. In choice $c_1$, for all events with $Q^2
< Q_0^2$, the parton distributions are estimated with $Q^2 = Q_0^2$,
while $\alpha_s$ and $x$ are kept at their true values. Hence this
choice overestimates the event rate. In choice $c_3$, all such events
are thrown away, so this choice underestimates the event rate.

The ``true'' cross-section would probably lie between the two extreme
estimates and is probably close to the cross-section with choice $c_2$,
where events with $Q^2 < 1$ GeV$^2$ are thrown away, while for events
with $1 <Q^2 < Q_0^2$, the parton distributions are computed with $Q^2$
rescaled to $Q_0^2$.

Notice that the uncertainties are larger for the mu case (and also will
be large for the electron case) and is less severe for the tau.

It should also be noted that this issue of perturbative tractability
cannot be cured by going to higher orders in the calculation. For
instance, the effect of including NLO contributions to the DIS
cross-section in the $c_1$ case in the limit of massless quarks and
target hadrons is also shown in Fig.~\ref{fig:dis}. The parton
distribution functions are again from the CTEQ6 NLO set while the
coefficient functions are given in Ref.~\cite{NLO}. While this is
correct for the muon case, there are further mass corrections for tau
interactions \cite{georgi,kretzer} which are severe since they lead to
effective higher twist terms. A discussion of these corrections can be
found in Ref.~\cite{kretzer}; however, the uncertainty from the
small-$Q^2$ events overwhelms the changes in going from LO to NLO. For
the numerical calculations that follow, therefore, we use the LO
expressions throughout, which are correct (including mass corrections)
for both muon and tau interactions.

Actual measurements at low and medium $Q^2$ are essential to close
this gap in our understanding of neutrino--nucleon
cross-sections. This may be possible in the near future with
measurements from MINOS, Minerva, T2K, Argoneut, etc. \cite{schmitz}.

\begin{figure}[htp]
\includegraphics[width=\textwidth,bb=18 144 592 440, clip]{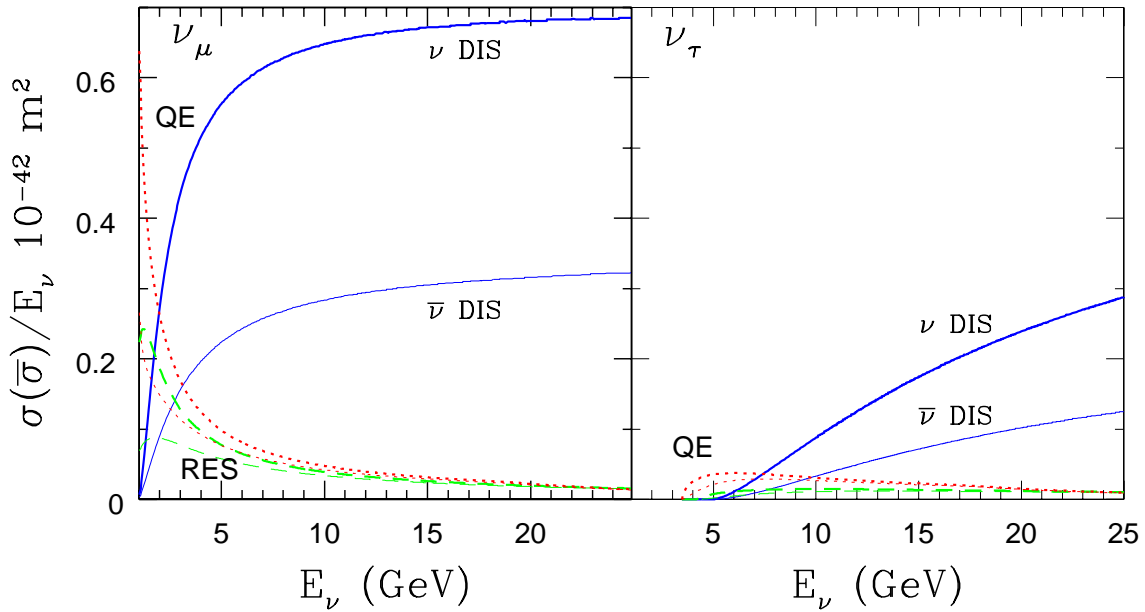}
\caption{Total quasi-elastic, resonance and deep inelastic scattering (DIS)
CC (anti)neutrino--nucleon cross-section as a function of the neutrino
energy for $\nu_\mu$ and $\nu_\tau$ interactions.}
\label{fig:cross}
\end{figure}

In summary, we use the CTEQ6 LO parton distribution set throughout
this paper, with choice $c_1$. The quasi-elastic, resonance and DIS
total cross-sections for $\nu_\mu$ and $\nu_\tau$ (and their
anti-particle)--nucleon CC interactions are shown in
Fig.~\ref{fig:cross}. The lower (upper) curve corresponds to
interaction of the (anti)-neutrino with an isoscalar nucleon (average
of proton and neutron). The actual numerical calculations use the
differential cross-sections shown in Eq.~(\ref{eq:cross}).

\subsection{Tau decay}

Here we focus on the leptonic decay mode of the tau, for example
$\tau^- \to \nu_\tau \mu^-\bar{\nu}_\mu$, where the
final muon from tau decay adds to the direct muon signal from
$\nu_\mu$--nucleon CC interactions. While the branching fraction into
muons (or electrons) is about 17\%, we are again interested in the
detailed kinematics of the final state muon and hence use the
differential decay rates; see Appendix A for details.

\begin{figure}[htp]
\includegraphics[angle=-90,width=0.8\textwidth]{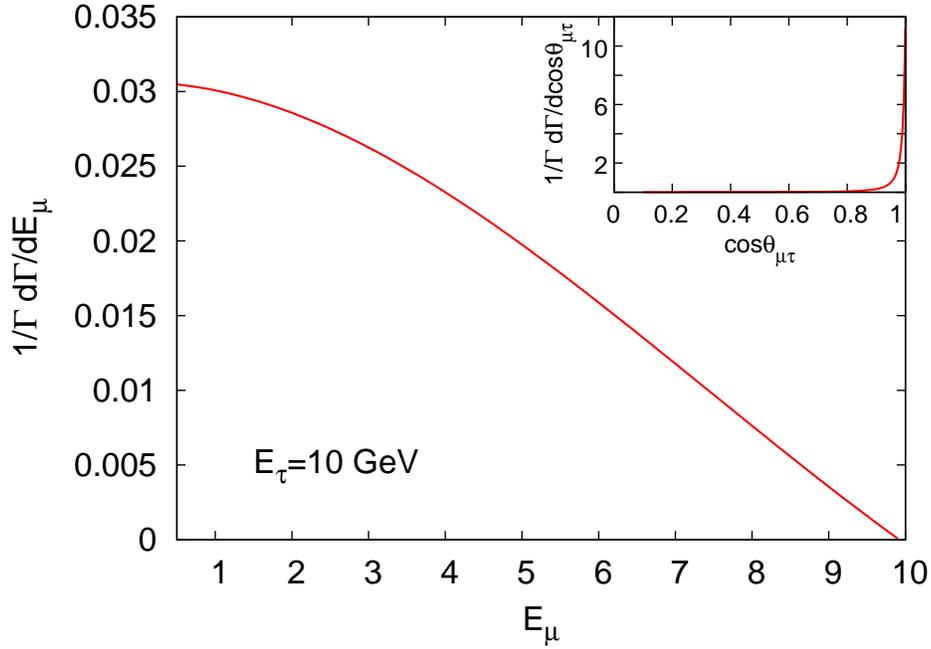}
\caption{Tau decay rate as a function of the final muon energy $E_\mu$ and
direction (inset) for tau with energy $E_\tau = 10$ GeV. Here $\theta_{\mu\tau}$ is
the opening angle between the muon and its parent tau.}
\label{fig:taudecay}
\end{figure}

The typical decay rate for a tau with energy $E_\tau=10$ GeV is shown as
a function of the muon energy and direction (with respect to that of
the parent tau) in Fig.~\ref{fig:taudecay}. Notice that the muon is
emitted mostly in the direction of the tau (which was itself forwardly
peaked) while its energy is typically highly degraded compared to its
parent. Hence, muons produced in $\nu_\tau$ interactions mostly contribute
to the {\em forward event rate at low energy}. The fractional decay rate
has been plotted so that the area under the curve in both cases is 17\%.

\section{Event rates in a far-detector}\label{sec:event}

\subsection{Preliminaries}\label{sec:prel}

We assume that the neutrinos interact with a 50 kton iron detector
such as the proposed INO/ICAL or MIND detector. Both $\mu^+$ and
$\mu^-$ beams are considered, with equal exposure for each. There are
two kinds of contributions in the signal:
\begin{enumerate}
\item The ``direct'' muons arising from CC interactions of muon
neutrinos (or anti-neutrinos) on the target.
\item The ``tau-induced'' muons arising from CC interactions of tau
neutrinos (or anti-neutrinos) on the target, with the subsequent decay
of the produced taus into muons. While the decay fraction is small,
the tau events are large, as mentioned earlier, and hence this contribution
is significant.
\end{enumerate}
In addition, there are right sign (RS) and wrong sign (WS) events for
each of these. Table~\ref{tab:rsws} clarifies the nature of different
sources of muons in the final state. The charge of the RS/WS lepton
depends on the charge of the muon in the storage ring. However, we do
not identify the charge of the final charged lepton: all muon events are
simply added. This is because the RS events are sensitive to deviations
from maximality of $\theta_{23}$; the WS events are very small compared to
the RS events and their inclusion may only marginally worsen the results;
however, the advantage gained in being ``charge-blind'' is significant,
especially in the errors.

\begin{table}[htp]
\begin{tabular}{|c|c|c|c|c|c|} \hline
Beam in & Neutrinos & \multicolumn{2}{c|}{RS muons} & \multicolumn{2}{c|}{WS muons} \\ 
Storage ring & in beam & Direct & Tau & Direct & Tau \\ \hline
\raisebox{-0.2cm}{$\mu^-$} & $\nu_\mu$ & $N_\mu \cdot P_{\mu\mu} \cdot
\sigma_\mu$ & $N_\mu \cdot P_{\mu\tau} \cdot \sigma_\tau \cdot \Gamma_\tau$ & -- & -- \\
& $\overline{\nu}_e$ & -- & -- &
$\overline{N}_e \cdot \overline{P}_{e\mu}\cdot \overline{\sigma}_\mu$ & 
$\overline{N}_e \cdot \overline{P}_{e\tau} \cdot \overline{\sigma}_\tau
\cdot \Gamma_\tau$ \\ \hline
\raisebox{-0.2cm}{$\mu^+$} & $\overline{\nu}_\mu$ &
$\overline{N}_\mu \cdot \overline{P}_{\mu\mu} \cdot \overline{\sigma}_\mu$ & 
$\overline{N}_\mu \cdot \overline{P}_{\mu\tau} \cdot 
\overline{\sigma}_\tau \cdot \Gamma_\tau$
& -- & -- \\
& $\nu_e$ & -- & -- &
$N_e \cdot P_{e\mu} \cdot \sigma_\mu$ & 
$N_e \cdot P_{e\tau} \cdot \sigma_\tau \cdot \Gamma_\tau$ \\ \hline
\end{tabular}
\caption{The terms contributing to the production of
right sign (RS) and wrong-sign (WS) muons from $\nu_e$ and $\nu_\mu$
neutrinos and anti-neutrinos with $\mu^\pm$ beams in the storage ring.
Negatively (positively) charged muons constitute the RS (WS) signal
with neutrinos from a $\mu^-$ storage ring and vice versa with $\mu^+$
beams. Event rates are proportional to the product of the oscillation
probabilities, $P_{ij}$, the neutrino fluxes, $N_{e,\mu}$, the differential
CC cross-sections, $\sigma_{\mu,\tau}$, and the differential decay rate,
$\Gamma_\tau$ (in the case of $\tau$-induced muon production), as is
symbolically shown.}
\label{tab:rsws}
\end{table}

The muon event rate is calculated as a function of the observed muon
energy (and not that of the original neutrino that participated
in the interaction). We have, generically, the number of muon events
in a detector at a distance L,
\begin{eqnarray}
{\cal{R}}^{b}_{i,D}(E) & = & K \int_{E_\nu^{\rm thr}}^{E_\nu^{\rm max}} \d E_\nu ~ \by{\d N_i(E_\nu,L)}{\d E_\nu}
\cdot P_{i\mu}(E_\nu,L) \int_0^1 \,\d\!\cos\theta_\mu \int_{E_-}^{E_+} \d E_\mu\by{\d\sigma_\mu (E_\nu,E_\mu,\theta_\mu)}
{\d E_\mu\d\cos\theta_\mu} \cdot R(E_\mu,E)~, \nonumber \\
{\cal{R}}^{b}_{i,\tau}(E) & = & K \int_{E_\nu^{\rm thr}}^{E_\nu^{\rm max}} \d E_\nu ~ \by{\d N_i(E_\nu,L)}{\d E_\nu}
\cdot P_{i\tau}(E_\nu,L) \int_{c_{\rm min}}^1\d\!\cos\theta_\tau \int_{E_-}^{E_+}\d E_\tau
\by{\d\sigma_\tau (E_\nu,E_\tau,\theta_\tau)}{\d E_\tau \d\cos\theta_\tau} 
\cdot \nonumber \\
 & & \hspace{3cm} \int_{\rm restr.}\d E_\mu \d\!\cos\theta_\mu \d\phi_\mu ~
\by{1}{\Gamma_\tau} \by{\d\Gamma(E_\tau, E_\mu, \theta_\tau, \theta_\mu,\phi_\mu)}{\d E_\mu
\d\cos\theta_\mu\d\phi_\mu} \cdot R(E_\mu,E)~,
\label{eq:rate} 
\end{eqnarray}
where the superscript $b$ refers to the beam-type, $\mu^-$ or $\mu^+$
circulating in the storage ring; $i = \mu,\bar{e} (\bar{\mu}, e)$
denotes the neutrino type produced in a $\mu^-(\mu^+)$ beam. Here
$K=N_t n_y$, $N_t$ is the number of target nucleons (assumed
isoscalar) and $n_y$ is the number of years of data accumulation. The
first expression in Eq.~(\ref{eq:rate}) corresponds to the production
of direct muon (denoted by the subscript \small{D}) production while
the second corresponds to $\tau$ production with subsequent decay into
muons (denoted by $\tau$), with $E$ being the observed energy of the
muons. $E_\nu^{\rm max}$ is the maximum neutrino energy in the factory
flux; the limits of integration are defined in Appendix A.
Note that the integration over the final muon
parameters, in case of tau production and decay, is restricted
by an angular constraint; see Appendix A for details. The charge-blind
events are obtained by summing over the index $i$ and hence the
total events for each beam are then obtained by the sum,
\begin{equation} {\cal{R}}^{b}_{tot}(E) = \sum_i
 ({\cal{R}}^{b}_{i,D}(E) + {\cal{R}}^{b}_{i,\tau}(E))~.
\label{eq:total-rate}
\end{equation}

Due to finite detector resolution, the observed lepton energy $E$
may not correspond to the true lepton energy. We assume that the muons
are reconstructed with a gaussian resolution width of 7\% ($\sigma =
0.07E_\mu$) for energetic, GeV muons. (We focus on the charged muon
distribution and do not try to reconstruct the energy and direction of
the original neutrino.) This identifies $R$ in Eq.~(\ref{eq:rate}) as the
Gaussian energy resolution function of width $\sigma$.

We assume a 90\% reconstruction efficiency of muons which is
charge-independent, although we implicitly assume an iron calorimeter
detector with charge identification capability as in the {\sc minos}
\cite{minos} and the proposed {\sc ical/ino} \cite{ino} or simulated
MIND detectors. We use the final muon (from direct muon production or via
tau decay) event rates accumulated over five years, to study the
sensitivity to the deviation of the mixing angle $\theta_{23}$ from
maximality.

\subsection{The event rates}
\begin{figure}[htp]
\includegraphics[width=\textwidth]{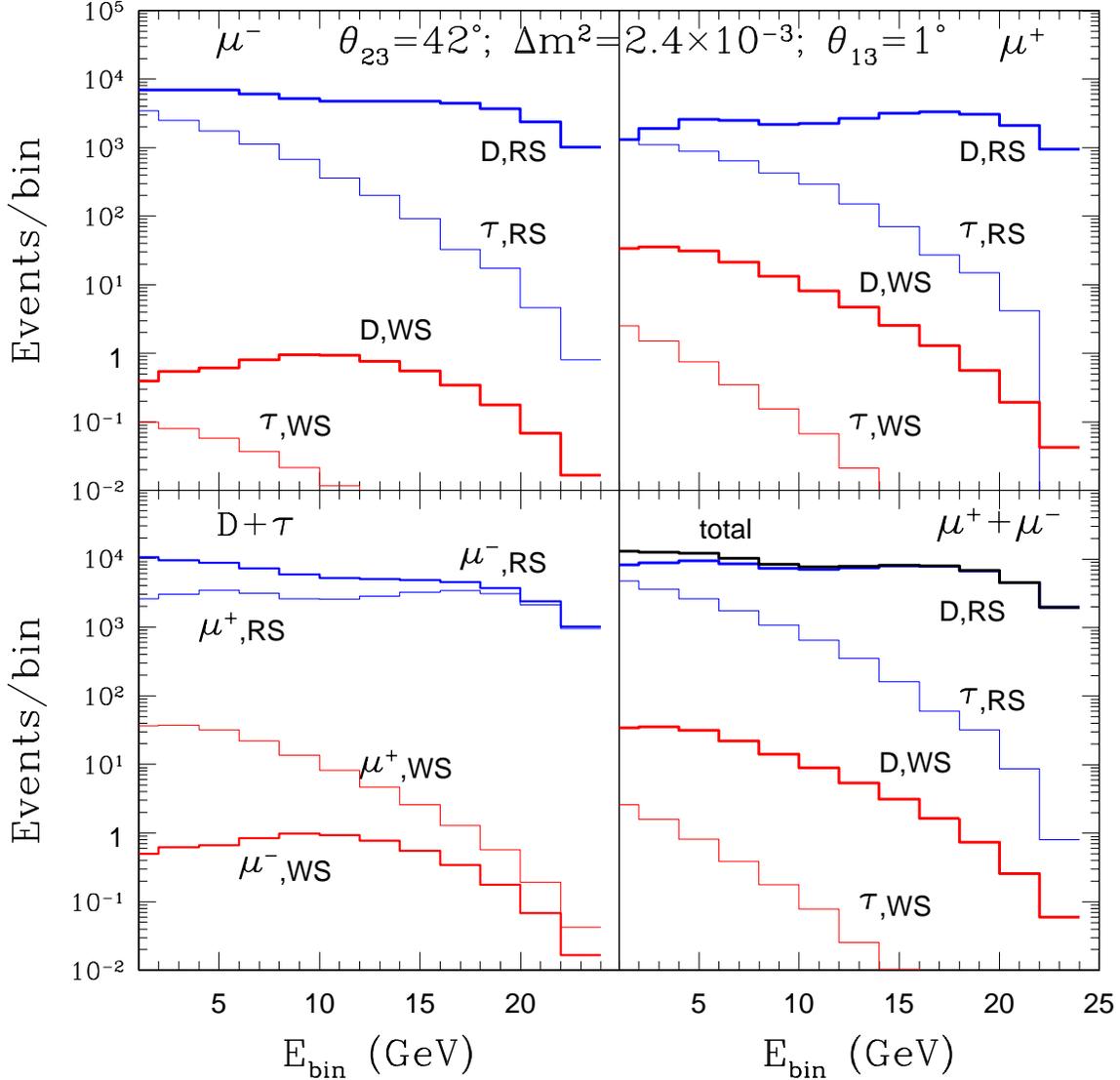}
\caption{Muon event rates over 5 years as a function of the observed
  muon energy, in bins of 2 GeV at a 50 kton muon detector distant
  $L=7,400$ km away from a muon factory neutrino source. Right sign
  (RS) and wrong sign (WS) events from $\mu^-$ and $\mu^+$ beams are
  shown in the upper panels. Contributions from direct muon production
  (denoted by $D$) and that of muons from tau decay (labeled as
  $\tau$) are separately shown. The left lower panel shows the sum 
  $D+\tau$. Notice that there is a substantial
  contribution from tau decay to RS events, as seen in the right lower
  panel; also, the WS events have a negligible contribution to the
  total event rate. Oscillation parameters are as shown, with $\Delta
  m^2$ in eV$^2$ and other oscillation parameters as given in
  Table~\ref{tab:bestfit}.}
\label{fig:data}
\end{figure}

At the magic \cite{magic} base-line, $L=7,400$ km, which is chosen
here, there is no sensitivity to the CP phase $\delta_{CP}$ which we
henceforth set (arbitrarily) to zero. Typical event rates for the
leading atmospheric parameters, $\Delta m^2=2.4\times 10^{-3}$ eV$^2$,
$\theta_{23}=42^\circ$ ($\sin^22\theta_{23}=0.9891$), the reactor
angle $\theta_{13}=1^\circ$ ($\sin^2\theta_{13}=0.0003$;
$\sin^22\theta_{13}=0.0012$) and the solar parameters set to the
central values given in Table~\ref{tab:bestfit}, are shown as a
function of the observed charged lepton energy in Fig.~\ref{fig:data}.
The panels show the direct muon production and tau decay contributions
to the right sign (RS) and wrong sign (WS) observed muon events from
$\mu^-$ and $\mu^+$ beams. Since the neutrino-nucleon cross-sections
are larger than those for anti-neutrinos, the $\mu^-$ ($\mu^+$) beams
have larger RS (WS) events. Hence, $\mu^+$ beams are preferred for
studying WS events (where charge identification is crucial) which are
sensitive to non-zero $\theta_{13}$, 23 mass hierarchy, octant of
$\theta_{23}$ and the CP phase $\delta$, away from the magic baseline.
A $\mu^-$ beam with larger RS events would be desirable for precision
of $\Delta m$ and $\theta_{23}$. However, using the full sample
without charge identification is preferable, as the detection
efficiency of the charge blind sample would be higher. Since the WS
events are a small fraction of the large RS sample, using the sum does
not worsen the precision of the parameters. We therefore add the
events (RS+WS) from both $\mu^-$ and $\mu^+$ beams; the total event
rate is also shown in the figure.

It can be seen that there is a substantial contribution to the RS
events from tau decay into muons. Since the tau life-time is very
short, these events will add to the observed RS events from direct
muon production in $\nu_\mu$--nucleon interactions. Since the tau
contribution alters the sensitivity to the oscillation parameters as
discussed in Section~\ref{sec:tau} and illustrated in
Fig.~\ref{fig:theta23}, we next discuss whether these events can be
removed through various cuts on the final charged lepton energy or
production angle.

\subsection{Cuts on tau contribution}

As stated earlier, tau production in neutrino--nucleon interactions is
extremely forward-peaked; see Fig.~\ref{fig:emX}. Furthermore, the
muons produced in leptonic tau decay are also forward peaked; see
Fig.~\ref{fig:taudecay}. Hence, one obvious way to remove/reduce the
tau contribution is an angular cut. Also, as can be seen from
Fig.~\ref{fig:taudecay}, the tau decay rate is large for muons
produced with low energies, resulting in a substantial tau
contribution to the muon events, at small observed muon energies (see
Fig.~\ref {fig:data}). Hence, a muon energy cut can also be
contemplated.

\begin{figure}[htp]
\includegraphics[width=\textwidth]{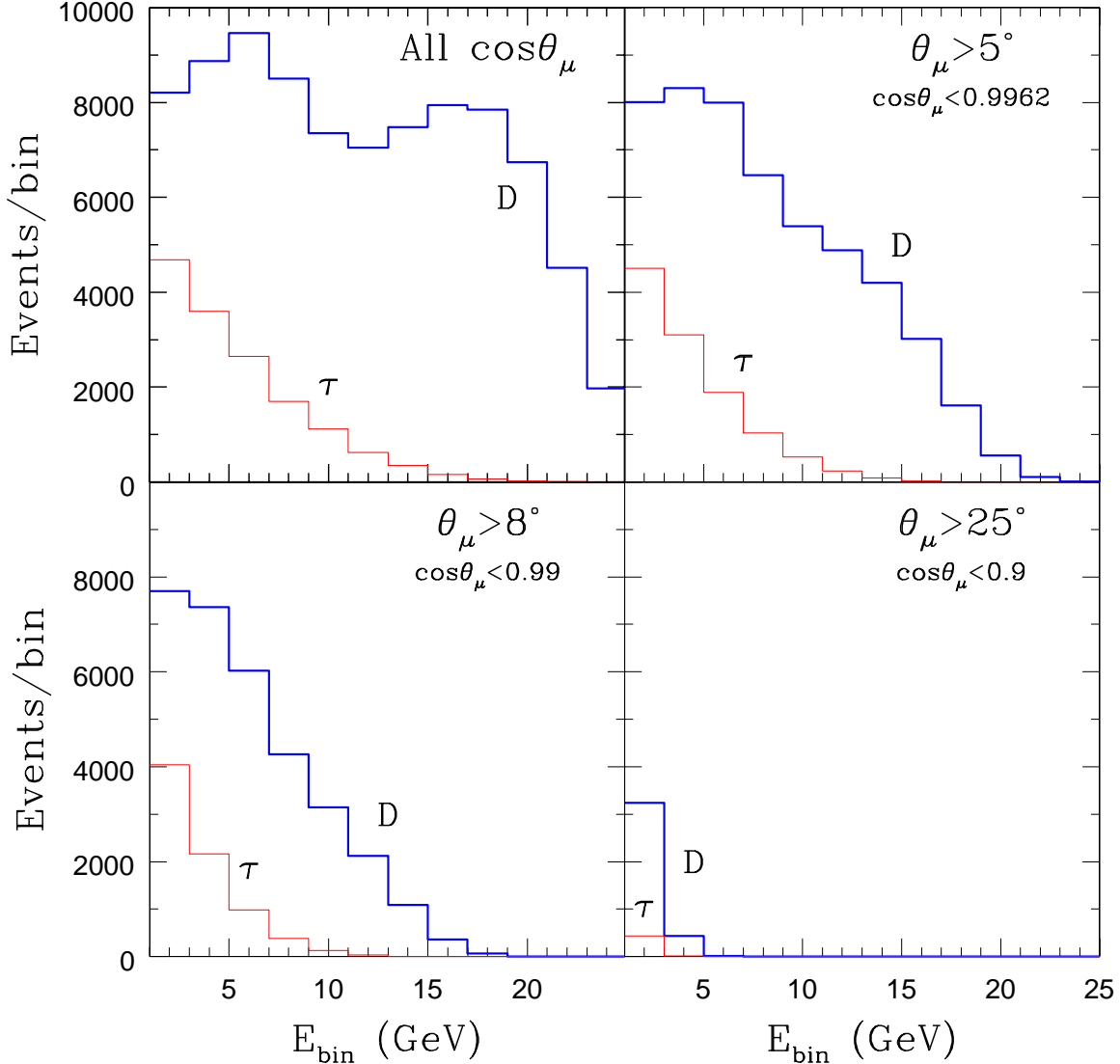}
\caption{Effects of angular cuts on the tau contribution to muon events
at neutrino factories. For details, see the text. It may not be possible
to devise effective cuts to remove the tau contribution which must
therefore be included in any analysis.}
\label{fig:cuts}
\end{figure}

The effect of cuts on the event rates is seen in Fig.~\ref{fig:cuts}.
The panels show the effect of an increasingly drastic angular cut on
the final state muons. It can be seen that the only cut effective in
removing/reducing the tau contribution is one (with $\theta_\mu >
25^\circ$) that removes the signal itself! Alternately a muon energy
cut, of about $E > 10$--15 GeV, can substantially remove the tau
contribution, still leaving sufficient direct muons. However, such a
large energy cut will will worsen the precision to which the mixing
parameters can be measured as sensitivity is higher in the lower
energy bins where matter effects are larger. In short, it is not
feasible to cut out the tau contribution and still make a precision
measurement, in this case, of the deviation of $\theta_{23}$ from
maximality.

\subsection{Effect of the tau contribution}

\begin{figure}[htp]
\includegraphics[width=\textwidth,bb=18 298 592 570, clip]{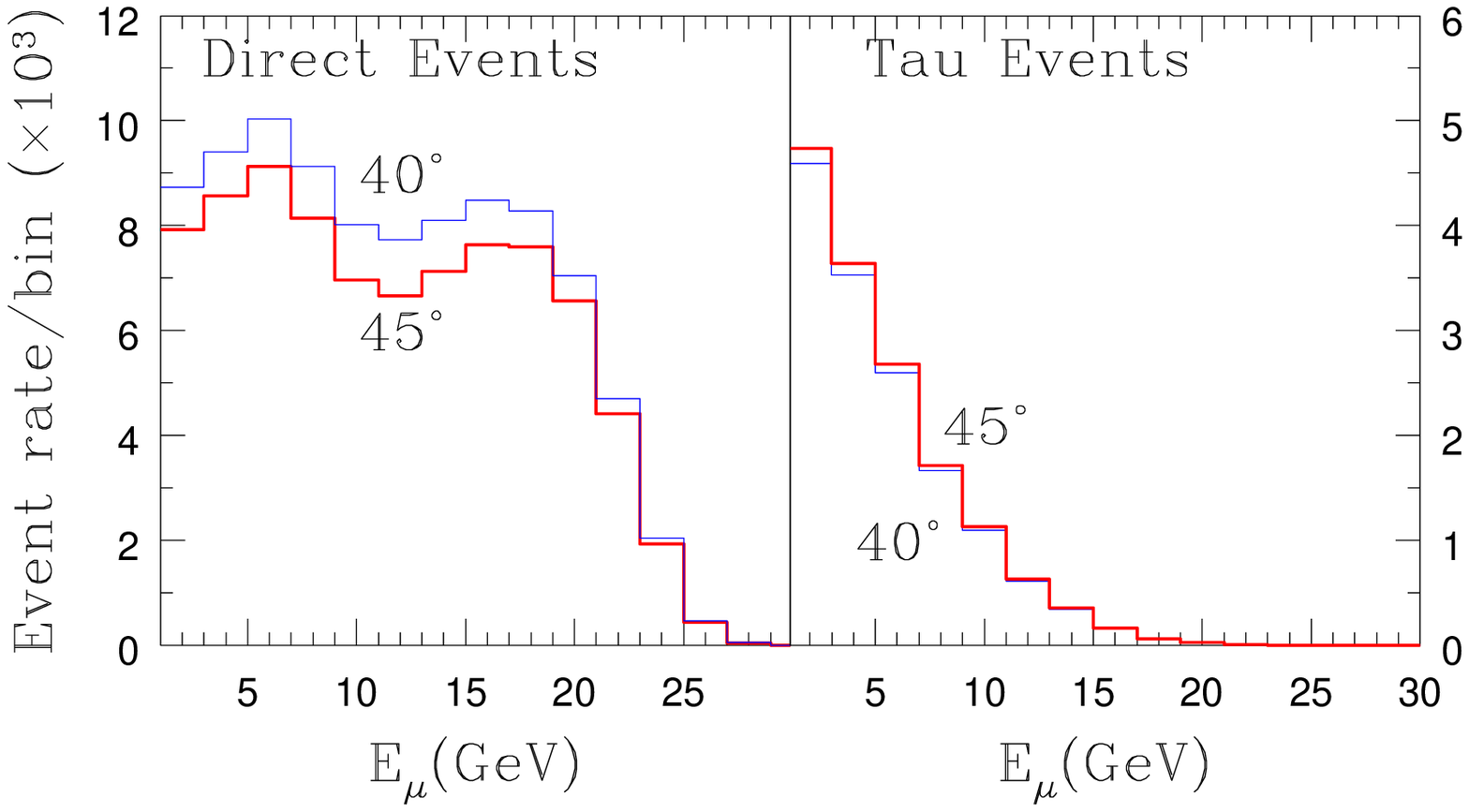}
\caption{Variation of direct and tau induced muon events with
$\theta_{23}$. While increase in $\theta_{23}$ decreases the direct
muon events, those coming from production and decay of taus tend to
increase marginally with larger $\theta_{23}$. The conflicting
behavior results in less sensitivity in the total events.
}
\label{fig:theta23}
\end{figure}

As stated earlier, the tau contribution alters the dependence on the
mixing parameters, and thus alters the precision to which we can
determine them. Fig.~\ref{fig:theta23} shows the dependence of the
direct and tau-induced muon events on $\theta_{23}$. While the tau
events have less sensitivity to this parameter, the rate increases
while the direct event rate decreases as $\theta_{23}$ increases
towards maximal $\theta_{23}=\pi/4$. However, the inclusion of muons
from tau events alters the uncertainties considerably.

A look back at the
expressions in Eq.~(\ref{eq:rate}) clearly shows a
major difference between direct and total (including tau contribution)
muon production that is most easily appreciated by explicitly writing
out the dominant RS contribution in each case:
\begin{eqnarray}
\d {\cal{R}}^b_D (RS) & \sim & P_{\mu\mu} \d N_\mu \d\sigma_\mu~; \nonumber \\
\d {\cal{R}}^b_\tau (RS) & \sim & P_{\mu\tau} \d N_\mu \d\sigma_\tau
\d\Gamma_\tau~.
\end{eqnarray}
A near detector (where the oscillation probabilities are near zero and
the survival probabilities are nearly 1), sensitive to muons, precisely
measures the combination $(\d N_\mu \d\sigma_\mu)$, that is, flux times
cross-section of the muons. This is what appears in the (RS) event rate
for direct muon production and is therefore very well constrained from
measurements at the near detector. Indeed, uncertainties are reduced to
a factor of $10^{-3}$, mainly arising from differences in the shape of
the flux at the near- and far-detectors. 

However, what appears in the (RS) event rate expression via tau
production and decay is the combination $\d N_\mu \d\sigma_\tau$.
While the flux is the same as for direct muon production and in fact
very well understood, the same is not true of the cross-section. While
the muon cross-sections have large uncertainties inherent in any
perturbative QCD calculation at low/medium $Q^2$, these are compounded for
the case of the heavy tau where mass corrections are large. Furthermore,
since these contributions result from oscillations, no near detector
can help reduce the uncertainties more. Hence overall uncertainties are
much larger for the tau contribution than for direct muons and this
exacerbates the problem of precision measurements with tau
contamination.

We now proceed to show this effect through specific numerical
calculations. We use a combined overall normalization error of 0.1\%
for direct muon events (separate flux and cross-section normalization
errors are not required as for WS studies since the RS events are by
far the dominant contribution and errors are kept well in check by
studies at the near detector as described above). We use a modest
normalization error of 2\% for the total (direct+tau) events; since
the same (muon neutrino) flux contributes, this is essentially the
error on the ratio of the tau to muon cross-sections. We use typical
input values of ($\Delta m^2, \theta_{23}, \theta_{13}$) to estimate
how well the generated ``data'' can be fitted, and calculate the
resulting precision on the parameters. Again, we keep the solar
parameters fixed at their best-fit values and set $\delta_{CP}$ to
zero. The best fits (and regions of confidence levels in parameter
space) are obtained by minimizing the chi-squared according to the
method of ``pulls'' \cite{pull} :
\begin{equation} 
\chi^2 = {\stackrel{{\hbox{min}}}{\xi}} \left[ \sum_{\rm bin=1}^N
\by{\left(\overline{\cal{R}}_{\rm
bin}^{th}(\xi) - {\cal{R}}_{\rm bin}^{data} \right)^2}
{\sigma_{\rm bin}^2} + \xi^2 \right]~. 
\label{eq:chi2}
\end{equation}
Here the normalization uncertainty for the event rate in a bin is
parameterized as a linear function of the pull:
\begin{equation}
\overline{\cal{R}}_{\rm bin}^{th}(\xi) = 
{\cal{R}}_{\rm bin}^{th}\left(1 + \Delta N \xi\right)~,
\end{equation}
where $\Delta N$ is either 0.1 or 2\% for direct and total respectively
and $\xi$ is the pull that accounts for the overall systematic error
on the rate. While the expression can be minimized over a set of pulls
for each systematic error, this is not needed since the rate is
dominated by the RS events as explained earlier. Here the theoretical
rates in a bin correspond to a central value of $\xi = 0$ with
$\xi=\pm 1$ representing $1\sigma$ errors. We follow the prescription
of Ref.~\cite{pull}, where the chi-squared is first
minimized over $\xi$ and then minimized with respect to oscillation
parameters to get the best-fit values and regions.

\subsection{Deviation of $\theta_{23}$ from maximality}

\begin{figure}[hbp]
\includegraphics[width=\textwidth,bb=18 178 592 672, clip]{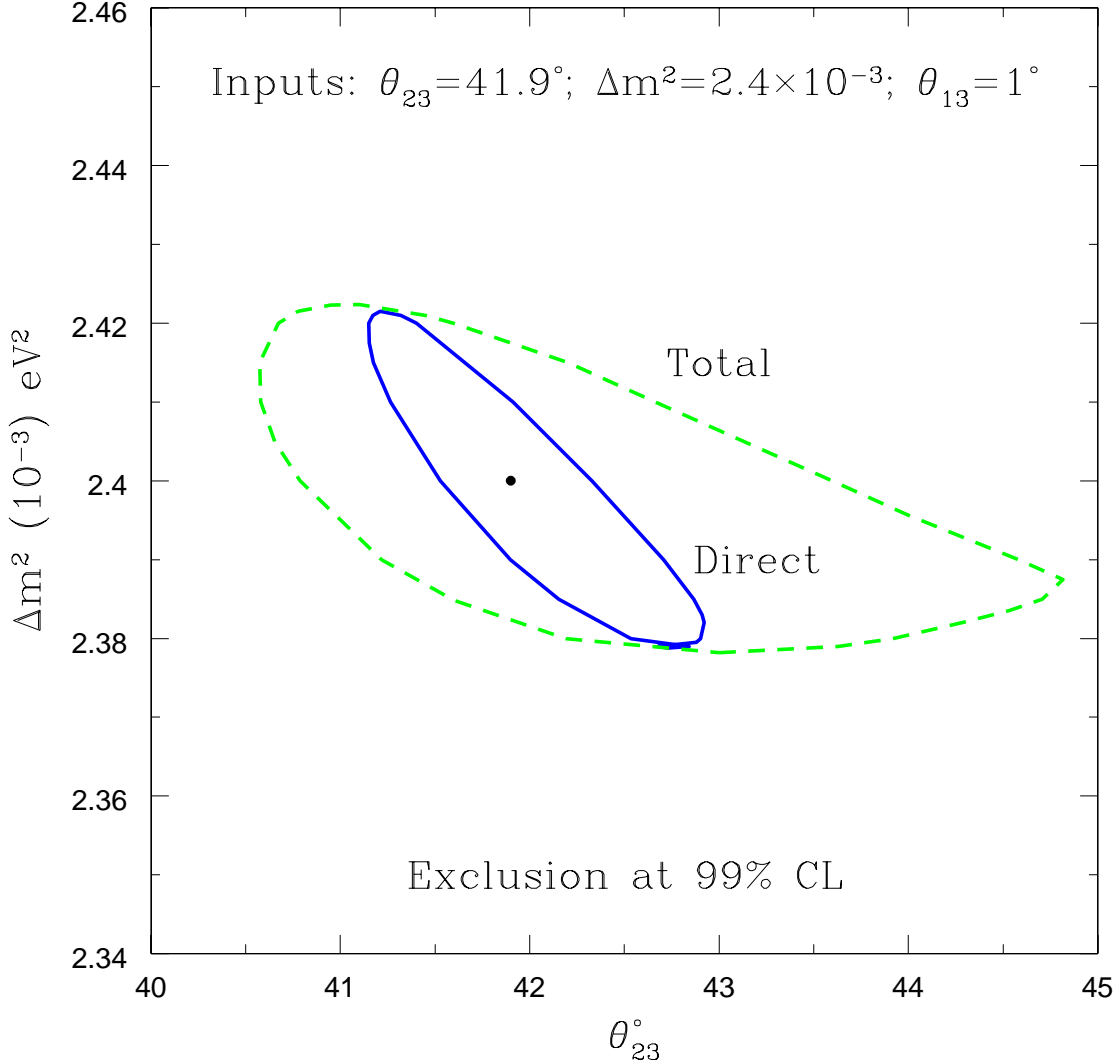}
\caption{Allowed parameter space in $\Delta m^2$--$\theta_{23}$ space at
99\% CL from CC events directly producing muons (solid line) and with
the inclusion of muons from tau decay as well (dashed line) for input
parameter values of 
($\Delta m^2, \theta_{23}, \theta_{13} = 2.4\times 10^{-3}$ eV$^2$,
$41.9^\circ$, $1^\circ$).} 
\label{fig:contt13fix}
\end{figure}

We present results for a typical sample set of input parameters,
($\Delta m^2, \theta_{23}, \theta_{13} = 2.4\times 10^{-3}$ eV$^2$,
$41.9^\circ$, $1^\circ$). After minimizing over the pull, we minimize over
$\Delta m^2$ and $\theta_{23}$, keeping the 13 mixing angle fixed.
Fig.~\ref{fig:contt13fix} shows the allowed region in $\Delta
m^2$--$\theta_{23}$ parameter space at 99\% CL ($\Delta \chi^2 =
9.21$). The solid line correspond to considering direct muon events
alone and the dashed line correspond to using the total events,
including those from tau decay.

Note that the selected signal (all muons in the final state,
independent of charge) is dominated by RS muons and is therefore
insensitive to the {\em octant} of $\theta_{23}$ as well as the {\em
sign} of $\Delta m^2$ (or the mass hierarchy). These can be probed
only through a study of WS muons, as has been discussed elsewhere.
Returning to the issue of deviation of $\theta_{23}$ from maximality,
it can be seen that the 99\% CL contour is much more constrained with
direct muons than for total muons, including those from tau decay. In
particular, it is the $\Delta m^2$ values that are smaller than the
input (true) value, that broaden the contour and limit the
discrimination. Hence the effect of adding in the tau contribution is
a worsening in the precision with which $\theta_{23}$ and its
deviation from maximality can be measured: a spread of $\sim 2^\circ$ if
tau events are removed and $\sim 4.5^\circ$ when they are included. This
occurs because of the conflicting dependence on $\theta_{23}$ of the
two contributions, as has been discussed earlier. However, it can be
seen from Fig.~\ref{fig:contt13fix} that the inclusion of the tau
events does not affect the determination of (the modulus of) $\Delta
m^2$ which is tightly constrained to better than $1\%$ in either case.

\begin{figure}[htp]
\includegraphics[width=\textwidth]{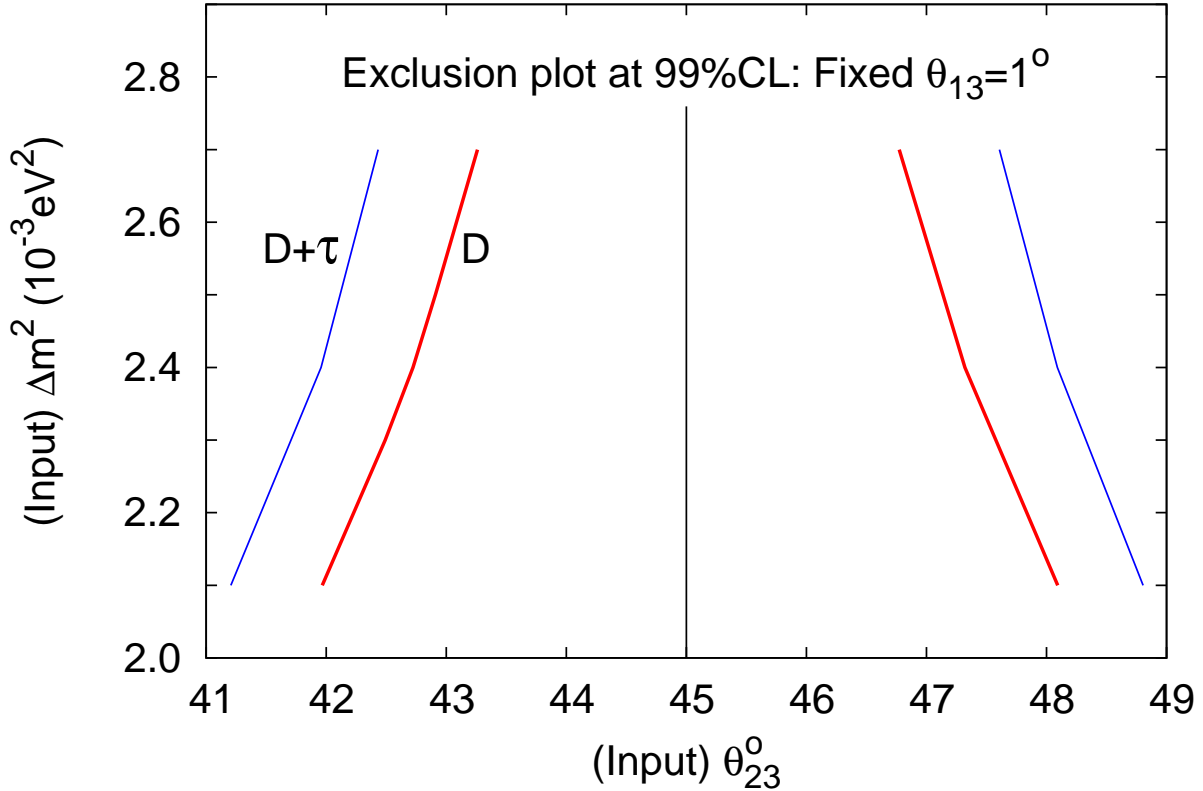}
\caption{The figure shows the largest (smallest) true value of
$\theta_{23}$ in the first (second) octant that can be discriminated
from a maximal value of $\theta_{23} = \pi/4$ as a function of $\Delta
m^2$ when only Direct ($D$) and total ($D + \tau$) events are considered.
Here $\theta_{13}$ is kept fixed: $\theta_{13} = 1^\circ$.}
\label{fig:disct13fix}
\end{figure}

The largest true value of $\theta_{23}$ that can be discriminated from
maximal through a study of muon events in neutrino factories is shown in
Fig.~\ref{fig:disct13fix}, as a function of $\Delta m^2$. Again,
$\theta_{13}$ is kept fixed at $\theta_{13} = 1^\circ$. While there is a
distinct but mild dependence on $\Delta m^2$, it is seen that tau
contamination worsens the ability to discriminate $\theta_{23}$ from
maximal, thus making this measurement as well as an octant measurement
harder than originally expected.

\subsection{Effect of $\theta_{13}$}

\begin{figure}[hbp]
\includegraphics[width=\textwidth,bb=18 390 592 680, clip]{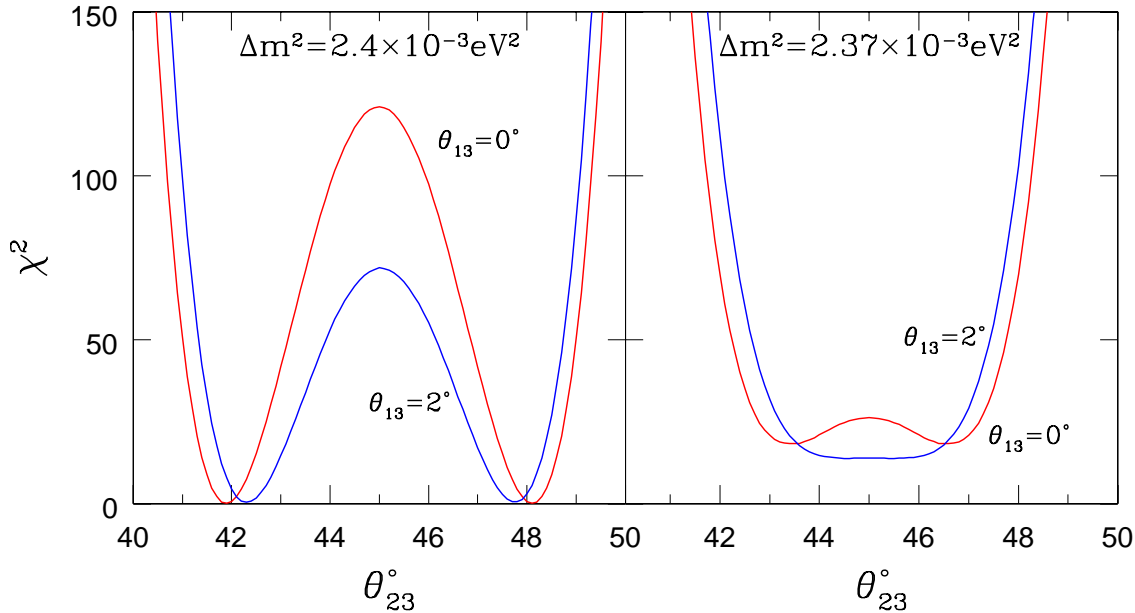}
\caption{Effect of $\theta_{13}$ on $\chi^2$ for input
parameter values of 
($\Delta m^2, \theta_{23}, \theta_{13} = 2.4\times 10^{-3}$ eV$^2$,
$41.9^\circ$, $1^\circ$). Increase in $\theta_{13}$ lowers the $\chi^2$
near maximality, and hence worsens the power of discriminating against
it, as shown for two typical values of $\Delta m^2$, for the same
input parameter values.}
\label{fig:chi2}
\end{figure}

Due to the extraordinarily large RS signal, it is anticipated that some
of the early measurements at a neutrino factory will be those of
precision determination of $\Delta m^2$ and $\theta_{23}$ and the
deviation of the latter from maximality. Certainly a strong case for
conventional neutrino factories of the type considered here will be made
only when $\theta_{13}$ is small, below the reach of near-future
super-beams or reactor searches. Hence we expect $\theta_{13}$ to be
relatively unknown at the time of such measurements, typically
restricted to a value $\theta_{13} < 2^\circ$ ($\sin^2 2\theta_{13} <
0.005$). The uncertainty in $\theta_{13}$ will worsen the allowed
parameter space from both direct and total muon events. In particular,
the dependence of the event rates (both direct and tau-induced) is such
that values of $\theta_{13}$ larger than the true value lower the $\chi^2$
near maximal $\theta_{23}$, and hence worsen the power of discriminating against
it. In other words, the tighter the upper bound on $\theta_{13}$, better
will be the discrimination of $\theta_{23}$ from maximality. This can
be seen from Fig.~\ref{fig:chi2} where the chi-squared is plotted (for
a fixed normalization, $\xi =0$, for clarity) for two different $\Delta m^2$
values, for the same input set as before.

\begin{figure}[hbp]
\includegraphics[width=\textwidth,bb=18 163 592 688, clip]{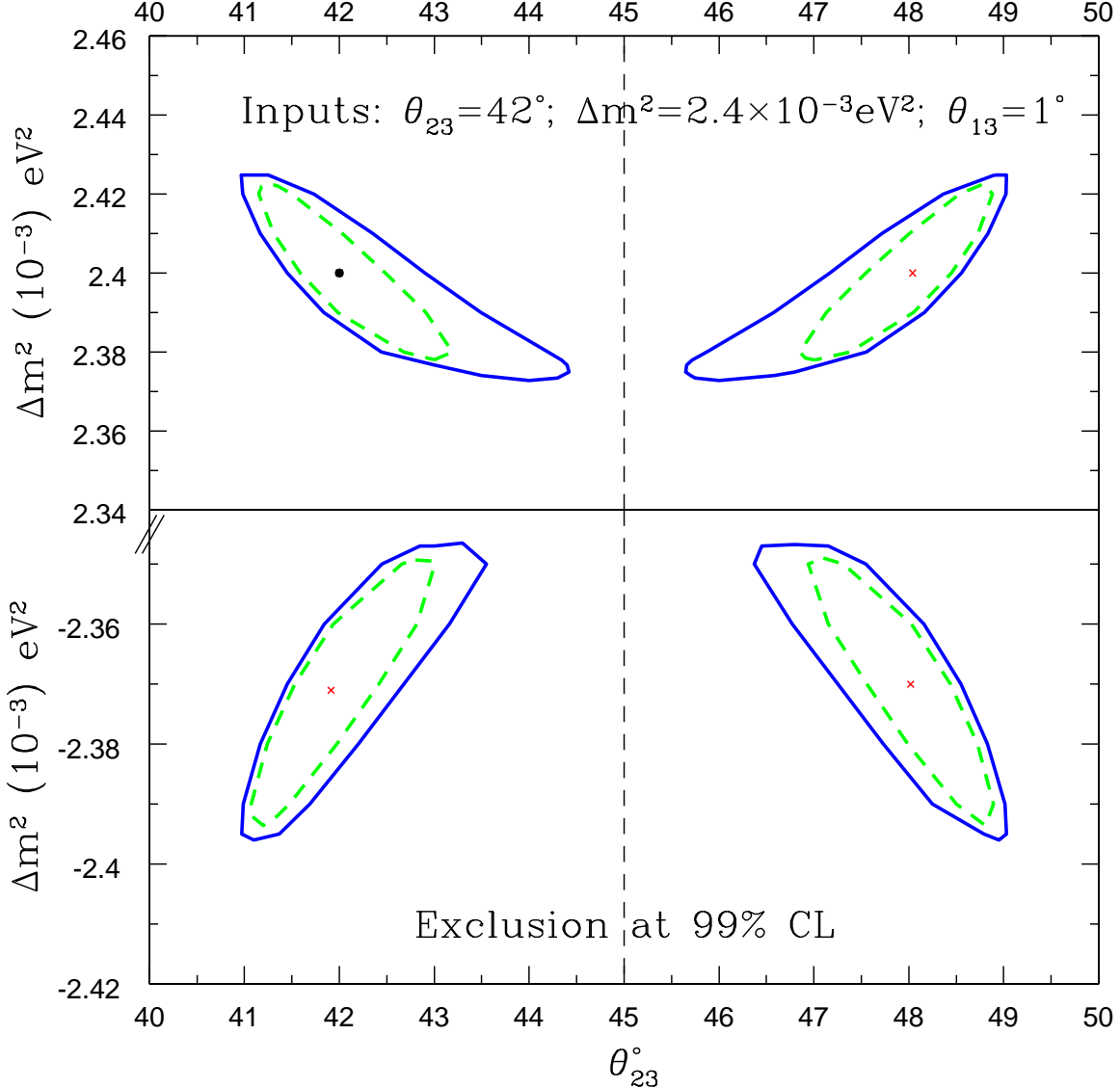}
\caption{Allowed parameter space in $\Delta m^2$--$\theta_{23}$ space at
99\% CL from all CC events producing muons, including tau events, with a
fixed normalization ($\xi=0$). The dashed
lines corresponds to fits with a fixed $\theta_{13}=1^\circ$ and the
solid lines result from varying $\theta_{13}$ with $0 < \theta_{13} <
2^\circ$. The input parameter values are
($\Delta m^2, \theta_{23}, \theta_{13} = 2.4\times 10^{-3}$ eV$^2$,
$42.0^\circ$, $1^\circ$). For completeness, the contours in the second
octant as well as with the inverted hierarchy are shown.} 
\label{fig:contnfix}
\end{figure}

The effect again leads to a widening of the contour in regions where
$\Delta m^2$ is smaller than the true value, as can be seen from
Fig.~\ref{fig:contnfix}. Here, the normalization is kept fixed to
$\xi=0$ for clarity in understanding the effect of varying
$\theta_{13}$ and we have assumed a bound $\theta_{13} < 2^\circ$,
with the contour corresponding to allowed parameter values at 99\% CL.
Fig.~\ref{fig:contnfix} also shows, for completeness, the false minima
in the second octant and in both octants with the inverted hierarchy.
Similar results are obtained if the true $\theta_{23}$ is in the
second octant and/or the hierarchy is inverted, although the precision
is marginally worse when the true $\theta_{23}$ is in the second
octant.

\begin{figure}[htp]
\includegraphics[width=\textwidth,bb=18 300 592 673, clip]{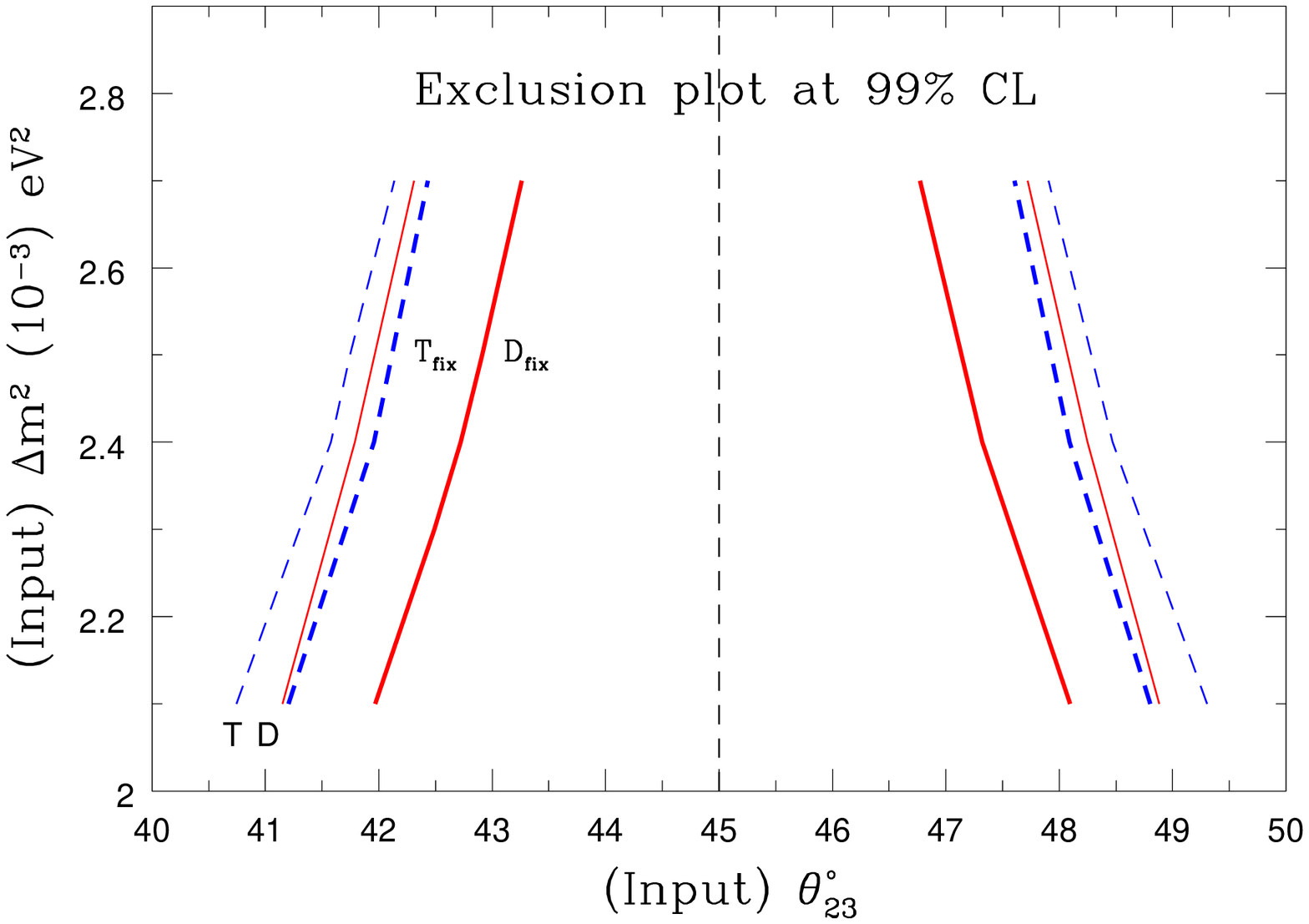}
\caption{The largest (smallest) true value of $\theta_{23}$ in the
first (second) octant that can be discriminated from a maximal value
of $\theta_{23} = \pi/4$ as a function of the true $\Delta m^2$ when
uncertainties in $\Delta m^2$, $\theta_{13}$ and the normalization
are taken into account. All the analyses use an input value of
$\theta_{13}=1^\circ$. The solid (dashed) lines are for direct muon
events only (total $=$ direct $+$ tau). The lighter obtained with
an uncertainty of $\sigma_{13} = 0.005$ on $\sin^22\theta_{13}$ will
approach the darker curves corresponding to fixed $\theta_{13}$ as the
uncertainty on $\theta_{13}$ decreases.}
\label{fig:disctot}
\end{figure}

Alternately, we can allow $\theta_{13}$ to vary freely, but include a
prior on this parameter, through an additional term in the
chi-squared to include the expected bound on $\sin^22\theta_{13}$
\cite{bound} from upcoming experiments,
\begin{equation}
\chi^2 \to \chi^2 + \by{\left( \sin^22\theta_{13} - \sin^22\theta_{13}^{\rm
true}\right)^2} { \sigma_{13}^2}~,
\end{equation}
where $\sigma_{13}$ is the error on $\sin^22\theta_{13}^{\rm true}$
which we take to be $\sigma_{13}=0.005$. Using this modified definition
of the chi-squared, and the same procedure as before, the 99\% CL
contours in the parameter space now correspond to $\Delta \chi^2 =
11.36$ for 3 parameters, $\Delta m^2$, $\theta_{23}$ and $\theta_{13}$.
In Fig.~\ref{fig:disctot} we show the largest true value of $\theta_{23}$
that can be discriminated from maximal when both $\Delta m^2$ and
$\theta_{13}$ are varied, as a function of the true $\Delta m^2$, given the
normalization uncertainties as specified earlier. The inclusion of
tau events still worsens the measurement, although the situation is not
as acute as when $\theta_{13}$ was kept fixed. Again, there is better
discrimination for larger $\Delta m^2$.

\paragraph{Some miscellaneous remarks}: We note that the ability to
discriminate from maximality is marginally asymmetric between the two
octants when $\theta_{13}$ is different from zero, as is the case in all
the examples shown. The results are better if the true value lies in the
first octant but not significantly so. Furthermore, the discrimination
capability improves as the uncertainty $\sigma_{13}$ decreases. Finally,
there is a non-trivial dependence of the results on the muon beam
energy, $E_b$. While the event rate increases with $E_b$, the number
of ``useful'' events does not, since the sensitivity to mixing
parameters is best for intermediate neutrino energies, $E_\nu \sim
5$--10 GeV. This arises from the energy dependence of the oscillation
probabilities as well as the energy dependence of Earth matter effects.
It appears that muon beams with energies 15--25 GeV are best suited for
such measurements.

\section{Conclusion}
We have re-examined the precision to which the leading atmospheric
parameters can be measured at a standard neutrino factory. We have
addressed the little-studied issue of contamination of the (right
or wrong sign) muon events sample due to oscillations of the muon or
electron neutrinos (anti-neutrinos) to tau neutrinos (anti-neutrinos).
These, through charge current interactions in the detector result in
tau leptons, which can then decay to muons, adding to the right as well
as wrong sign muon events obtained directly, without tau production.
The tau contribution {\em worsens} the precision to which $\theta_{23}$
can be measured while leaving the determination of the atmospheric
mass squared difference $\vert \Delta m^2 \vert$ unchanged. Hence the
tau contamination worsens the ability to discriminate against maximal
$\nu_\mu \leftrightarrow \nu_\tau$ mixing. Any cuts imposed to remove the
tau events drastically reduce the events from direct muon production as
well and are hence impracticable. Uncertainties from this tau background
have therefore first to be brought under control before making precision
parameter measurements at neutrino factories.

\section*{Appendix A}

We detail the kinematics of neutrino--nucleus CC interactions with muons
in the final state (direct muon events) and that of CC interactions
with tau in the final state with subsequent tau decay into muons in the
lab frame (tau-induced muon events). The events are characterized by the
final muon energy and direction, i.e., the entire muon spectrum. Hence the
detailed kinematics of the interactions, both for direct muon production
and via tau decay, are required.

We define the four-momenta of the incoming neutrino ($k$), target
nucleon ($p$), produced $\tau$ lepton ($k^\prime$) and the muon
($q_1$) coming from the decay of the tau in the laboratory frame as,
\begin{eqnarray} \nonumber
k & = & (E_\nu,0,0,E_\nu)~, \nonumber \\
p & = & (M,0,0,0)~, \nonumber \\
k^\prime & = & (E_\tau, p_\tau \sin\theta_\tau,0, p_\tau
\cos\theta_\tau)~, \nonumber \\
q_1 & = & (E_\mu, p_\mu \sin\theta_\mu\cos\phi_\mu, p_\mu
\sin\theta_\mu\sin\phi_\mu, p_\mu\cos\theta_\mu)~,
\label{eq:momenta}
\end{eqnarray}
where $M$ is the (isoscalar) nucleon mass. The kinematics is standard
and straightforward for the case of $\nu_\mu$ interactions (where there
is no tau production, all components of $k^\prime$ are zero), with the
angles of the direct muon produced in the CC interaction taking values
$0 < \cos\theta_\mu < 1$ and $\phi_\mu =0$.

The threshold for the interaction is given by $E_\nu^{\rm thr} =
m_l+m_l^2/2M$, for $l=\mu, \tau$; hence it is significant, $E_\nu \gtrsim
3.5$ GeV, in the case of tau production. Furthermore, in the case of
$\nu_\tau$ interactions, the tau is produced at a very forward angle,
$c_{\rm min} < \cos\theta_\tau < 1$, in the lab frame, where
\begin{equation}
c_{\rm min} (E_\nu) = \sqrt{1+\by{M}{E_\nu} + \by{M^2}{E_\nu^2}
- \by{m_\tau^2}{4E_\nu^2} - \by{M^2}{m_\tau^2}}~.
\label{eq:cmin}
\end{equation}
Finally, the angle $\phi_\mu$ of the muon produced during tau decay is
restricted by the decay kinematics to obey the constraint:
\begin{equation}
\cos\phi_\mu > \by{2(E_\mu E_\tau - p_\mu p_\tau \cos\theta_\tau
  \cos\theta_\mu)-(m_\tau^2+m_\mu^2)}{2 p_\mu p_\tau \sin\theta_\tau
  \sin\theta_\mu}~.
\label{eq:cphi}  
\end{equation}
In addition, the available phase space in both direct muon production
and tau-induced muon production, restricts the available lepton energy
to $E_-< E_l < E_+$, where $l= \mu, \tau$. We have,
\begin{equation}
E_{\pm} (E_\nu, \cos\theta_l) = \by{1}{2a} (b\pm \sqrt{b^2-4ac})~,
\end{equation}
with 
\begin{eqnarray} \nonumber
a & = & (E_\nu+M)^2 - E_\nu^2 \cos^2\theta_l~, \nonumber \\
b & = & (E_\nu+M) (2M E_\nu + m_l^2)~, \nonumber \\
c & = & m_l^2 E_\nu^2 \cos^2\theta_l + (M E_\nu + m_l^2/2)^2~.
\end{eqnarray}
At the limits, $m_X = m_{X,{\rm min}}=M$, the nucleon mass. The notation is
standard: $m_X^2 = W^2 = (p+q)^2$, where $q=k^\prime-k$ is the 4-momentum
of the intermediate gauge boson in the lab frame; $m_X$ is constrained
to lie in the range $\sqrt s-m_l \ge m_X \ge M $, with $s=(k+p)^2$ .

The expression for semi-leptonic decay, $\tau \to \mu e \nu_e$, of a tau
into a final state muon, with four-momenta $k^\prime$ and $q_1$
respectively (defined in Eq.~(\ref{eq:momenta})), in the lab frame of the
original neutrino-nucleon interaction, is given by,
\begin{eqnarray}
\by{\d\Gamma (E_\tau,E_\mu,\theta_\tau,\theta_\mu,\phi_\mu)}{\d E_\mu
\d\cos\theta_\mu\d\phi_\mu} & = & G_F^2
\by{p_\mu}{48 E_\tau \pi^4} \left[3(m_\tau^2+m_\mu^2) (k^\prime\cdot q_1)
- 4(k^\prime\cdot q_1)^2-2m_\tau^2 m_\mu^2\right ]~.
\end{eqnarray}
The total decay width of the tau is $\Gamma_\tau = 1/c\tau$
where $c\tau = E_\tau/m_\tau (c\tau_0)$ where the rest-frame lifetime is
$c\tau_0 = 87.2$ microns.

\vspace{0.5cm}

\noindent {\bf Acknowledgments}: We thank M.V.N. Murthy for discussions and
a careful reading of the manuscript. One of us (NS) thanks Murthy for
suggesting a study of precision measurements at neutrino factories.

\end{document}